\documentclass[fleqn,usenatbib]{mnras}

\usepackage{newtxtext,newtxmath}

\usepackage[T1]{fontenc}

\DeclareRobustCommand{\VAN}[3]{#2}
\let\VANthebibliography\thebibliography
\def\thebibliography{\DeclareRobustCommand{\VAN}[3]{##3}\VANthebibliography}

\usepackage{graphicx}	
\usepackage{amsmath, bm}	% Advanced maths commands
\usepackage{amssymb}	% Extra maths symbols
\usepackage{mydefs}     % place my own commands
\usepackage[dvipsnames]{xcolor}     % colorful comments 
\usepackage{float} %required for the placement specifier H
\usepackage{ulem}
\usepackage{xspace}

\title[BAO forecast for CSST]{Fisher forecast for the BAO measurements from the CSST spectroscopic and photometric galaxy clustering}

\author[Z. Ding et al.]{
Zhejie Ding,$^{1,2}$\thanks{zd585612@ohio.edu}
Yu Yu,$^{1,2}$\thanks{yuyu22@sjtu.edu.cn}
Pengjie Zhang$^{1,2,3}$
\\
$^{1}$Department of Astronomy, School of Physics and Astronomy, Shanghai Jiao Tong University, Shanghai 200240, China\\
$^{2}$Key Laboratory for Particle Astrophysics and Cosmology (MOE)/Shanghai Key Laboratory for Particle Physics and Cosmology, Shanghai 200240, China\\
$^{3}$Tsung-Dao Lee Institute, Shanghai Jiao Tong University, Shanghai 200240, China
}

\date{Accepted XXX. Received YYY; in original form ZZZ}

\pubyear{2023}

\begin{document}
\label{firstpage}
\pagerange{\pageref{firstpage}--\pageref{lastpage}}
\maketitle

\begin{abstract}
The China Space Station Telescope (CSST) is a forthcoming Stage IV galaxy survey. It will simultaneously undertake the photometric redshift (photo-z) and slitless spectroscopic redshift (spec-z) surveys mainly for weak lensing and galaxy clustering studies. The two surveys cover the same sky area and overlap on the redshift range. At $z>1$, due to the sparse number density of the spec-z sample, it limits the constraints on the scale of baryon acoustic oscillations (BAO).
By cross-correlating the spec-z sample with the high density photo-z sample, we can effectively enhance the constraints on the angular diameter distances $D_A(z)$ from the BAO measurement.
Based on the Fisher matrix, we forecast a $\geq$ 30 per cent improvement on constraining $D_A(z)$ from the joint analysis of the spec-z and cross galaxy power spectra at $1.0<z<1.2$.
Such improvement is generally robust against different systematic effects including the systematic noise and the redshift success rate of the spec-z survey, as well as the photo-z error. We also show the BAO constraints from other Stage-IV spectroscopic surveys for the comparison with CSST.
Our study can be a reference for the future BAO analysis on real CSST data. The methodology can be applied to other surveys with spec-z and photo-z data in the same survey volume.
\end{abstract}

\begin{keywords}
cosmology: theory -- distance scale -- large-scale structure of Universe
\end{keywords}

%%%%%%%%%%%%%%%%%%%%%%%%%%%%%%%%%%%%%%%%%%%%%%%%%%

%%%%%%%%%%%%%%%%% BODY OF PAPER %%%%%%%%%%%%%%%%%%

\section{Introduction}
In modern cosmology, it is mostly puzzling to understand the nature of dark energy. 
In the framework of Einstein's general theory of relativity, dark energy was introduced to explain the accelerated expansion of the Universe, which was firstly discovered by measuring the luminosity distances of Type Ia supernovae (SNe Ia) \citep{Riess1998, Perlmutter1999}. 
Apart from SNe Ia, the scale of baryon acoustic oscillations (BAO) in galaxy clustering is another primary probe to measure the cosmic expansion rate \citep[e.g. see the review of][]{Weinberg2004}. 
BAO are the sound waves generated from the initial density fluctuations at the early stage of the Universe. Due to the coupling between photons and baryons, the sound waves could propagate in the Universe until the recombination epoch at redshift $z\simeq 1100$, when photons decoupled from baryons taking away the radiation pressure. The largest distance that the sound waves could propagate is called the sound horizon, with the comoving size about $150$ Mpc \citep{Peebles1970, Sunyaev1970, Bond1984}. The sound horizon scale has been precisely measured from the cosmic microwave background \citep[CMB; e.g.][]{Hinshaw2013, Planck2018}. The BAO signature is also imprinted in the large-scale structure formed in the later Universe, and we can measure BAO statistically from the galaxy clustering.
Given the high-precision sound horizon scale measured from CMB, we can take it as a standard ruler and calibrate the BAO scale measured from galaxy clustering at different redshifts, in order to obtain the cosmological distances and cosmic expansion history. 

The last two decades have seen a series of large-scale spectroscopic redshift (spec-z) surveys conducted to map the three-dimensional distribution of galaxies, including the 2dF Galaxy Redshift Survey \citep{Cole2005}, the WiggleZ Dark Energy Survey \citep{Blake2011b}, the 6dF Redshift Survey \citep{Beutler2011}, as well as the prestigious Sloan Digital Sky Survey (SDSS) with different stages, e.g. the Baryon Oscillation Spectroscopic Survey (BOSS) of SDSS-III \citep[e.g.][]{Dawson2013, Anderson2014, Alam2017, Beutler2017}, and the completed SDSS-IV extended Baryon Oscillation Spectroscopic Survey \citep[e.g.][]{Ata2018, Ross2020, Neveux2020, Wang2020, Zhao2021, Alam2021}. Since the first BAO detection from galaxy clustering \citep{Eisenstein2005, Cole2005}, the precision of the BAO scale measurements has increased from 5 per cent to $1$ per cent level for redshifts $z<0.75$ \citep{Anderson2014, Beutler2017, Alam2017}. Furthermore, the measurements have been extended to higher redshifts and from different tracers, e.g. luminous red galaxies \citep[LRGs;][]{GilMarin2020, Bautista2021}, emission line galaxies \citep[ELGs;][]{Raichoor2021, Tamone2020}, quasars \citep{Hou2021, Neveux2020}, and Lyman alpha (Ly $\alpha$) forests \citep{duMasdesBourboux2020}.

In near future, there will be several Stage IV spectroscopy surveys, including the Prime Focus Spectrograph \citep[PFS;][]{Takada2014}, the Euclid \citep{Euclid2011}, and the Nancy Grace Roman Space Telescope \citep[hearafter Roman;][]{Spergel2015}. The Dark Energy Spectroscopic Instrument \citep[DESI;][]{DESI2016, Adame2023b, DESI2022} is the first Stage IV survey which has started the observation. With larger survey volume and galaxy number density, they will dramatically increase the constraints on cosmological parameters.
   
As one of the Stage IV galaxy surveys, CSST is a space-based telescope on the same orbit of the Chinese Manned Space Station \citep{Zhan2011,Zhan2018,Zhan2021, Gong2019}. It is planned to be launched around 2024. CSST is a 2-m telescope with a large field of view, i.e. $1.1\times 1.0$ deg$^2$, and will cover a total sky area 17500 deg$^2$ from the $10$-yr survey. As two main goals, it will perform the photometric imaging survey for billions of galaxies to probe weak gravitational lensing. Simultaneously, using the slitless spectroscopy, it will measure redshifts of millions of galaxies to study galaxy clustering. The redshift range spans $0 - 4.0$ and $0 - 2.5$ for the photo-z and spec-z surveys, respectively. Recently, \cite{Gong2019} predicted the constraints on the cosmological parameters from the CSST weak lensing (WL) and galaxy clustering statistics, and found a significant improvement from the joint analyses of WL, galaxy clustering and galaxy-galaxy lensing observables. As following studies, \cite{Miao2023} estimated the constraints on the cosmological and systematic parameters from individual probes or multiprobe of the CSST surveys. \cite{Lin2022} gave forecast on the sum of the neutrino mass constrained from the photo-z galaxy clustering and cosmic shear signal. 

In our study, we specifically focus on the BAO scale measurement from the CSST spec-z and photo-z galaxy clustering and their joint analyses. 
Not only from spec-z surveys, the BAO signal has been detected from multiple photo-z surveys \citep[e.g.][]{Padmanabhan2007, Estrada2009, Carnero2012, Seo2012, Sridhar2020, Abbott2019, Abbott2022, Chan2022}. Due to the large redshift error in photo-z surveys, it smears information along the line of sight. The BAO scale measurements from photo-z surveys can constrain the angular diameter distance $\DA$ relatively well, but not the Hubble parameter $H(z)$. However, a photo-z survey is more efficient to detect galaxies at higher redshifts, cover a larger sky area, and obtain a larger galaxy number density.
While for a spec-z survey, the BAO constraints can be quickly deteriorated as redshift goes higher due to the decreasing number density. 
It turns out that cross-correlating a sparse spec-z sample with a dense photo-z sample can effectively improve the constraints on $\DA$ compared to that from the spec-z tracer alone \citep{Nishizawa2013, Patej2018, Zarrouk2021}. Such benefit comes from the cancellation of the cosmic variance since both samples trace the underlying dark matter field in the same survey volume \citep{Eriksen2015}, which is the case for CSST.   
From the BAO measurement, we forecast the constraints on $\DA$ and $H(z)$ at different redshifts. We focus on the improvement from the joint analyses of the spec-z and photo-z clustering. Our study is complementary to the previous work on the forecast for the cosmological parameters, and can be a reference for the BAO detection from real data analysis.

This paper is structured as follows. 
In Section \ref{sec:csst_survey}, we give a brief summary of the CSST photo-z and spec-z surveys, and show the corresponding mock galaxy redshift distributions that we adopt. In Section \ref{sec:method}, we overview the methodology of the Fisher matrix. We show the BAO modelling in the galaxy auto and cross power spectra. We discuss the numerical setting in the Fisher forecast. In Section \ref{sec:result}, we show the Fisher forecasts of $\DA/\rd$ and $H(z)\rd$ from the spec-z, photo-z and their joint analyses. We study the systematic influence from the spec-z systematic noise, the spec-z redshift success rate, and the photo-z error, respectively. Finally, we conclude in Section \ref{sec:conclusions}.   
Throughout this paper, we use the flat lambda code dark matter ($\Lambda$CDM) cosmology based on \cite{Planck2016}, i.e. $\Omega_{\text{b}}h^2=0.0223$, $\Omega_{\text{c}}h^2=0.1188$, $n_s=0.9667$, $\sigma_8=0.816$, and $h=0.6774$. The value of magnitude is based on the AB system.

\section{CSST surveys}\label{sec:csst_survey}
The CSST will conduct the photo-z and spec-z surveys concurrently, covering a wide and overlapped sky area.
We summarize some instrumental parameters of the two surveys, and discuss the mock galaxy redshift distributions that we adopt for the analyses.  

\subsection{CSST photo-z survey}
The CSST photo-z imaging survey will use seven broad-band filters, i.e. NUV, $u$, $g$, $r$, $i$, $z$, and $y$ to cover the ultraviolet and visible light with the wavelength range 255--1000 nm \citep{Gong2019, Liu2023}. There will be four exposures for the NUV and $y$ bands, and two exposures for the other bands. Each exposure takes 150 s. For extended sources (galaxies), the magnitude limit of the $g$, $r$ and $i$ bands is $\sim 25$ mag, and the imaging resolution can reach $\sim 0.15$ arcsec \citep{Liu2023}.

The mock photo-$z$ redshift distribution is based on \cite{Cao2018} (hereafter Cao2018) that utilized the COSMOS galaxy catalogue \citep{Capak2007,Ilbert2009}. The COSMOS has a 2 deg$^2$ field, and covers a wide redshift range $0<z<5$ \citep{Ilbert2009}. By selecting the samples with $i^+ \leq 25.2$ and removing stars, X-ray and masked sources, Cao2018 obtained a cleaned catalogue. Taking the redshifts of the cleaned COSMOS catalogue as the true redshifts (input), Cao2018 measured the photo-$z$ using the spectral energy distribution (SED) template-fitting technique \citep[e.g.][]{Bruzual2003}. Furthermore, they selected sub-sets with different photo-z accuracy which is quantified by the normalized median absolute deviation $\sigma_{\text{NMAD}}$ \citep[e.g.][]{Ilbert2006,Brammer2008}, i.e. 
$\sigma_{\text{NMAD}} = 1.48\times \text{Median}\left(|\Delta z - \text{Median}(\Delta z)|/(1+z_s)\right)$,
where $\Delta z=z_{\text{s}} - z_{\text{p}}$, $z_{\text{s}}$ and $z_{\text{p}}$ denote the spec-z (or true redshift) and photo-z, respectively. The advantage of $\sigma_{\text{NMAD}}$ is not sensitive to the catastrophic redshift failure from the SED fitting. 
Meanwhile, it can represent the standard deviation of a Gaussian distribution.  

\begin{figure}
    \centering
    \includegraphics[width=\linewidth]{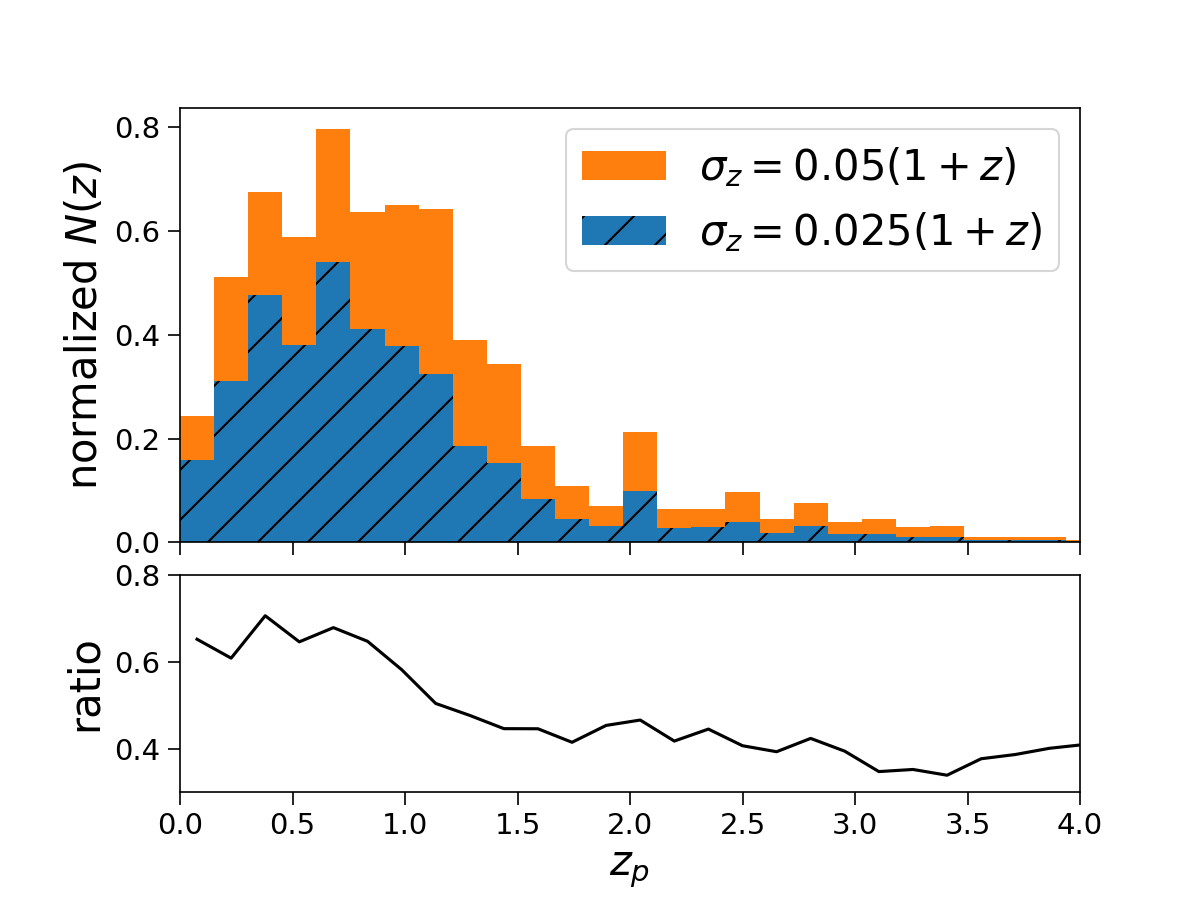}
    \caption{Normalized galaxy redshift distribution of the CSST photo-z survey. The mock distribution is built from the COSMOS catalogue. In the upper panel, the histogram with slashes denotes the sub-set of the COSMOS catalogue with the redshift error $\sigma_z=0.025(1+z)$. The other one denotes the sub-set with $\sigma_z=0.05(1+z)$. In the lower panel, we show the ratio between the two sub-sets.}
    \label{fig:nz_phot}
\end{figure}

Cao2018 selected about $95$ per cent and $58$ per cent of the overall cleaned sample and obtained $\sigma_{\text{NMAD}}\sim 0.05$ and $0.025$, respectively. 
In the upper panel of Fig. \ref{fig:nz_phot}, 
we show the normalized photo-z distribution with $\sigma_{\text{NMAD}}=0.05$. The distribution of $\sigma_{\text{NMAD}}=0.025$ (shown as the histogram with slashes inside) is rescaled by the ratio of the total galaxy numbers of the two distributions. In the lower panel, we show the galaxy number ratio from each bin with the bin width $\Delta z=0.15$. We cut redshift at $4.0$, beyond which the number density is low. In our default analyses, we ignore the effect from the photo-z outliers, and naively consider the root mean square (rms) of $z_p$ as
$\sigma_{\zp}=\sigma_{\text{NMAD}}(1+z_{\text{s}})$. 

\subsection{CSST spec-z survey}
\begin{figure}
    \centering
    \includegraphics[width=\linewidth]{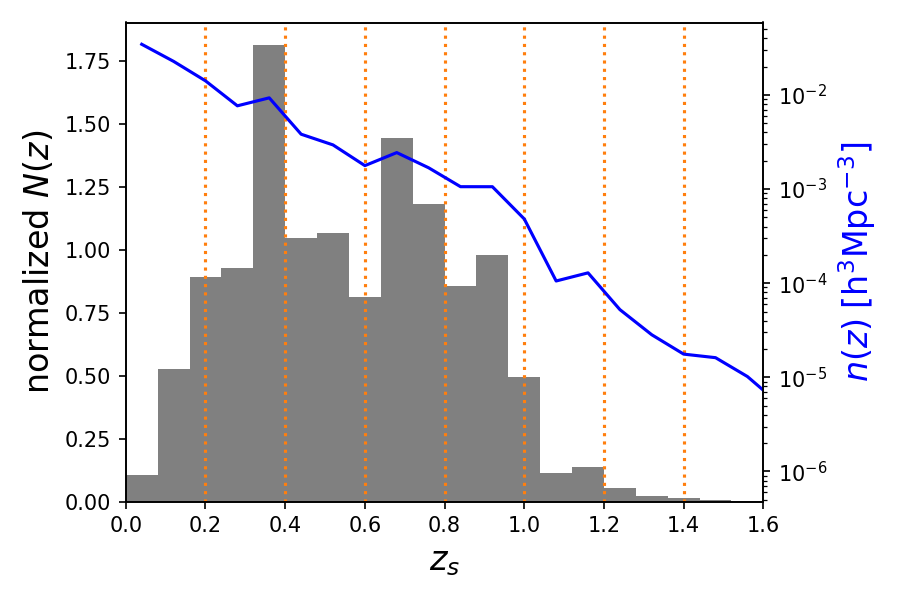}
    \caption{Galaxy redshift distribution of CSST slitless spec-z survey after normalization. We model the mock distribution based on the zCOSMOS catalogue. The histogram shows the normalized galaxy number distribution. The solid line denotes the galaxy volume number density distribution with the fiducial setting. The label on the right vertical axis represents the number density value. The density drops below $10^{-4}\hMpccube$ at $z>1.0$. The vertical dotted lines denote the boundaries of the eight redshift bins that we divide.}
    \label{fig:nz_spec}
\end{figure}

The CSST spec-z survey will use slitless gratings to measure spectroscopic redshifts. It has three bands GU, GV, and GI with the same wavelength range as that of the photo-z bands. The expected spectral resolution of each band is $R=\Delta \lambda/\lambda \geq 200$ \citep{Gong2019}. Following the method in \cite{Gong2019}, we construct the mock spec-z distribution based on the zCOSMOS catalogue \citep{Lilly2007, Lilly2009}, which contains 20690 galaxies in a 1.7 deg$^2$ field. The magnitude limit of the zCOSMOS is $I_{\text{AB}}=22.5$, comparable to that of the CSST spec-z survey \citep{Gong2019, Liu2023}. We select a sub-set of the zCOSMOS samples with high confidence on the galaxy redshift accuracy.\footnote{
We select samples with the confidence classes equal to 1.5, 2.4, 2.5, 9.3, 9.5, 3.x and 4.x as noted in the table 1 of \cite{Lilly2009}, apart from the mask on redshift ($z > 0$) and magnitude ($I_{\text{AB}}\leq 22.5$).} The sub-set contains about $80$ per cent of the total, and mainly distributes in the redshift range $0 < z \leq 1.5$. In Fig. \ref{fig:nz_spec}, we show the normalized spec-z distribution as the histogram. The galaxy number drops quickly beyond $z=1.0$. In addition, due to the relatively low spectroscopic resolution of the CSST slitless spectroscopy, we should not ignore the redshift error when we model the spec-z galaxy clustering signal. For our default setting, we adopt the spec-z error from \cite{Gong2019}, i.e. $\sigma_{\zs}=0.002(1+\zs)$, along with the redshift success rate of $f_{\text{eff}}^z=f^0_{\text{eff}}/(1+\zs)$, i.e. the fraction of galaxies reaching such redshift accuracy \citep{Wang2010}. $f^0_{\text{eff}}$ is the value at $\zs=0$. We adopt the moderate expectation of $f^0_{\text{eff}}$ as $0.5$ for the fiducial case, and show the comoving volume number density as the solid line in Fig. \ref{fig:nz_spec}.

\begin{table*}
	\centering
	\caption{Parameters of the CSST spec-z and photo-z surveys with the sky coverage $17500$ $\text{deg}^2$. We set eight tomographic bins in the redshift range $0<z<1.6$ with the bin width $0.2$. The galaxy bias is assumed as $b_g(z)=1+0.84z$ for both the spec-z and photo-z galaxy distributions. We show the number density of galaxies from the two surveys. For the spec-z sample, the redshift error is set to be $\sigma_z=0.002(1+z)$, and the number density is down-sampled by $0.5/(1+z)$ on the original distribution based on zCOSMOS. For the photo-z sample, we use the distribution with the redshift error $\sigma_z=0.025(1+z)$, and show the results in parentheses. The power spectrum damping parameter $\Sigma_z=c\sigma_z/H(z)$ due to the redshift measurement error is also given. Furthermore, we show the S/N ratio $\bar{n}_g P_g$ and the effective volume $V_{\text{eff}}$ at $(k=0.16\hMpc, \mu=0.6)$ and $(k=0.2\hMpc, \mu=0)$, respectively. As the default case, we assume the spec-z systematic noise $\Psys=0$.}
	\label{tab:survey_params}
	\begin{tabular}{c c c c c c c c c c c}
		\hline
		Redshift & $V_{\text{survey}}$ & $\bar{n}_g 10^4$ & Bias & $\Sigma_z$ & $\bar{n}_g P_g(0.16, 0.6)$ & $\bar{n}_g P_g(0.2, 0)$ & $V_{\text{eff}}(0.16, 0.6)$ & $V_{\text{eff}}(0.2, 0)$\\
		         & [$\Gpchcube$] & [$\hMpccube$] & $b_g$ & [$\Mpch$] &  & & [$\Gpchcube$] & [$\Gpchcube$] \\
		\hline
0.0 < z < 0.2 & 0.33 & 181.97 (1956.02) & 1.08 & 6.28 (78.54) & 35.706 (0) & 37.314 (401.095) & 0.313 (0) & 0.314 (0.329)\\
0.2 < z < 0.4 & 1.93 & 89.64 (857.31) & 1.25 & 6.66 (83.30) & 18.998 (0) & 20.172 (192.922) & 1.743 (0) & 1.753 (1.911)\\
0.4 < z < 0.6 & 4.23 & 26.52 (354.63) & 1.42 & 6.84 (85.47) & 5.496 (0) & 6.189 (82.748) & 3.025 (0) & 3.132 (4.125)\\
0.6 < z < 0.8 & 6.57 & 18.94 (283.44) & 1.59 & 6.87 (85.84) & 3.982 (0) & 4.557 (68.192) & 4.196 (0) & 4.416 (6.379)\\
0.8 < z < 1.0 & 8.64 & 8.95 (208.56) & 1.76 & 6.81 (85.08) & 1.898 (0) & 2.192 (51.084) & 3.705 (0) & 4.072 (8.307)\\
1.0 < z < 1.2 & 10.32 & 1.36 (113.72) & 1.92 & 6.69 (83.65) & 0.290 (0) & 0.337 (28.136) & 0.523 (0) & 0.655 (9.622)\\
1.2 < z < 1.4 & 11.61 & 0.21 (54.50) & 2.09 & 6.55 (81.86) & 0.045 (0) & 0.053 (13.555) & 0.022 (0) & 0.029 (10.073)\\
1.4 < z < 1.6 & 12.57 & 0.04 (39.96) & 2.26 & 6.39 (79.92) & 0.008 (0) & 0.009 (9.961) & 0.001 (0) & 0.001 (10.379)\\		
		\hline
	\end{tabular}
\end{table*}

Table \ref{tab:survey_params} shows the relevant parameters from the CSST photo-z and spec-z surveys for this work. We divide the redshift range $0<z<1.6$ into eight uniform bins. For each bin, we calculate the survey volume and galaxy number density, given the survey area $17500$ $\text{deg}^2$ and the galaxy redshift distributions. In addition, we show the galaxy bias, the galaxy power spectrum damping parameter from redshift error, the signal-to-noise (S/N) ratio, and the effective volume, respectively. We discuss how we set these parameters in Section \ref{sec:method}. The numbers in the parentheses denote the parameters for the photo-z survey.       

\section{Methodology}\label{sec:method}
In this study, we use the Fisher matrix formalism \citep{Fisher1935, Vogeley1996, Tegmark1997a} to forecast the constraints on the BAO scale from the CSST photo-z and spec-z galaxy clustering, as well as from their joint analyses.

\subsection{Fisher matrix of galaxy surveys}
For a galaxy survey, we can consider the galaxy power spectrum as the observable, denoted as $P(\bm{k})$, i.e. 
\begin{align}
\langle \delta(\bm{k})\delta(\bm{k'}) \rangle \equiv (2\pi)^3 \delta_\text{D}(\bm{k}+\bm{k'}) P(\bm{k}), \label{eq:pkmu}
\end{align}
where $\delta(\bm{k})$ is the galaxy number density fluctuation as a function of wave vector $\bm{k}$, $\langle \cdots \rangle$ denotes the ensemble average, and $\delta_{\text{D}}$ is the Dirac delta function.
Assuming the likelihood of the power spectrum as Gaussian distributed, the Fisher matrix can be expressed as \citep{Tegmark1997b, Seo2003} 
\begin{align}\label{eq:fisher_single_tracer}
\text{F}_{ij} = \int_{-1}^{1}\int_{k_{\text{min}}}^{k_{\text{max}}} \frac{2\pi k^2 dk d\mu}{2(2\pi)^3} V_{\text{survey}}\left[\frac{\partial\,P^T(k,\mu)}{\partial \theta_i} \text{C}^{-1}\frac{\partial\, P(k,\mu)}{\partial \theta_j}\right],
\end{align}
where $\mu$ is the cosine angle between the wave vector $\bm{k}$ and the line of sight, $V_{\text{survey}}$ is the survey volume, C denotes the covariance matrix of $P(k, \mu)$, the superscript $T$ denotes the transpose which is for $P(k, \mu)$ as an array in the joint analyses case (Eq. \ref{eq:P_multi}), and $\theta_i$ is the $i$th parameter of $P(k, \mu)$. The covariance matrix of the parameters can be calculated from the inverse of the Fisher matrix,
\begin{align}
\text{Cov}(\theta_i,\,\theta_j) = (F^{-1})_{ij}. 
\end{align}
The square root of each diagonal term of $\text{Cov}(\theta_i,\,\theta_j)$ gives the standard deviation of each parameter after marginalizing over the other parameters. In this study, we take the marginalized error as the Fisher forecast for the parameter constraint.

Under the Gaussian assumption, the inverse covariance matrix of the power spectrum only depends on the observed band power, i.e.  
\begin{align}
    \text{C}^{-1} = \left(\frac{1}{P(k, \mu)+1/\bar{n}_g}\right)^2,
\end{align}
where $1/\bar{n}_g$ is the Poisson shot noise, and $\bar{n}_g$ is the mean galaxy number density in a given redshift bin. Eq. (\ref{eq:fisher_single_tracer}) can also be expressed as 
\begin{align}
   \text{F}_{ij} = \int_{-1}^{1}\int_{k_{\text{min}}}^{k_{\text{max}}} \frac{2\pi k^2 dk d\mu}{2(2\pi)^3} V_{\text{eff}}\frac{\partial\,\text{ln} P^T(k,\mu)}{\partial \theta_i} \frac{\partial\, \text{ln} P(k,\mu)}{\partial \theta_j}, 
\end{align}
where $V_{\text{eff}}$ is the effective volume of the survey \citep{Feldman1994},
\begin{align}
    V_{\text{eff}} = \left(\frac{\bar{n}_g P(k, \mu)}{\bar{n}_g P(k, \mu) + 1}\right)^2 V_{\text{survey}},
\end{align}
which absorbs the S/N ratio, i.e. $\bar{n}_g P(k,\mu)$. If $\bar{n}_g P(k,\mu)\gg 1$, it is considered as the cosmic variance dominating, otherwise, it is the shot noise dominating. We show $\bar{n}_g P(k,\mu)$ at $(k=0.16\hMpc, \mu=0.6)$ and $(k=0.2\hMpc, \mu=0)$ as in \cite{DESI2016}, and the corresponding $V_{\text{eff}}$ in the right columns of Table \ref{tab:survey_params}. 

 In our study, we use the Fisher matrix format of Eq. (\ref{eq:fisher_single_tracer}) and extend it to the multitracer case, i.e. we are interested in the constraints from the joint analyses of the CSST spec-z, photo-z galaxy clustering and their cross-correlations. We modify the observable in Eq. (\ref{eq:fisher_single_tracer}) to be
\begin{align}\label{eq:P_multi}
P(k, \mu)=\big[
P_{\text{s}},\; 
P_{\text{sp}}, \;
P_{\text{p}} \big]^T,
\end{align}
where $P_{\text{s}}$ and $P_{\text{p}}$ denote the spec-z and photo-z galaxy power spectra, respectively. $P_{\text{sp}}$ is the cross power spectrum between the spec-z and photo-z data. We describe the modelling of the power spectra in the following sections. For the multitracer analyses, we need to consider the cross-correlations between the power spectra in the covariance matrix. We model the Gaussian covariance matrix C as \citep[e.g.][]{White2009, Zhao2016}
\begin{align}
\text{C}=
\begin{bmatrix} 
\hat{P}_{\text{s}}^2 & \hat{P}_{\text{s}} P_{\text{sp}} & P_{\text{sp}}^2 \\
\hat{P}_{\text{s}} P_{\text{sp}} & \frac{1}{2}(P_{\text{sp}}^2+\hat{P}_{\text{s}} \hat{P}_{\text{p}}) & \hat{P}_{\text{p}} P_{\text{sp}} \\
P_{\text{sp}}^2 & \hat{P}_{\text{p}} P_{\text{sp}} & \hat{P}_{\text{p}}^2
\end{bmatrix},\label{eq:cov_multitracer}
\end{align}
where the hat sign denotes the power spectrum including the shot noise. We ignore the shot noise in the cross power spectrum. In the case if we only consider a sub-set of $P$ in Eq. (\ref{eq:P_multi}), e.g. the joint analyses of the spec-z and cross power spectra, we have the observable and covariance as

\[
P(k, \mu)=
\begin{bmatrix}
P_{\text{s}} \\
P_{\text{sp}}
\end{bmatrix},\;
\text{C}=
\begin{bmatrix}
\hat{P}_{\text{s}}^2 & \hat{P}_{\text{s}} P_{\text{sp}} \\
\hat{P}_{\text{s}} P_{\text{sp}} & \frac{1}{2}(P_{\text{sp}}^2+\hat{P}_{\text{s}} \hat{P}_{\text{p}})
\end{bmatrix}.
\]
With only one component of Eq. (\ref{eq:P_multi}), the Fisher formalism reduces to the single tracer case as Eq. (\ref{eq:fisher_single_tracer}).

\subsection{BAO modelling}\label{sec:bao_model}
The anisotropic galaxy power spectrum in redshift space can be modelled phenomenologically \citep[e.g.][]{Beutler2017,Blanchard2020}, 
\begin{align}\label{eq:pkmu_bao}
    P_{g\text{, spec-z}} = \left[D(a) F_{\text{RSD}} F_{\text{zerr}} \right]^2\left[P_{\text{BAO, nl}}(k,\mu) + P_{\text{m, sm}}(k)\right] + P_{\text{sys}},
\end{align}
where 
$D(a)$ is the linear growth function depending on the scale factor $a\equiv (1+z)^{-1}$, and
\begin{align}
    F_{\text{RSD}} = (b_g+ f \mu^2) F_{\text{FoG}},
\end{align}
which is contributed from the redshift space distortions. It consists of two parts. One is the Kaiser effect shown as the bracket term \citep{Kaiser1987}, which boosts the clustering amplitude along the line of sight at large scales. $b_g$ is the linear galaxy bias, and $f$ is the linear growth rate of structure, defined as the logarithmic derivative of the growth function over the scale factor, i.e. $f\equiv d\,\text{ln}\,D/d\,\text{ln}\,a$. The other part is the Finger of God (FoG) effect due to the halo velocity dispersion,
which is widely adopted as the Lorentz form \citep[e.g.][]{Cole1995, Beutler2017}, i.e.
\begin{align}
    F_{\text{FoG}} = \frac{1}{1+k^2\mu^2\Sigma_{\text{FoG}}^2/2},
\end{align}
where $\Sigma_{\text{FoG}}$ is the damping parameter. Due to the difficulty of modelling FoG precisely, e.g. \cite{Ross2017} and \cite{Wang2017} simply set $\Sigma_{\text{FoG}}=4\Mpch$ in the analysis of the twelfth data release of BOSS over the redshift range $0.2<z<0.75$. In our analysis, we adopt a redshift dependent value of $\Sigma_{\text{FoG}}(z)=7/(1+z)\Mpch$ following \citet{Gong2019}.

The term $F_{\text{zerr}}$ models the damping on power spectrum due to the galaxy redshift measurement error. Since the redshift error applies a Gaussian kernel on the line-of-sight distance, the damping on the power spectrum is a Gaussian form too \citep[e.g.][]{Peacock1994, Seo2003}, i.e.
\begin{align}
  F_{\text{zerr}} = \exp(-k^2\mu^2\Sigma_z^2/2),
\end{align}
with the damping parameter 
\begin{align}
    \Sigma_z = \frac{c\sigma_z}{H(z)},\label{eq:Sigma_z}
\end{align}
where $c$ is the speed of light, $H(z)$ is the Hubble parameter, and $\sigma_z$ is the redshift uncertainty. If $\sigma_z$ is small, e.g. as in SDSS and DESI with fibre spectroscopy, such term can be ignored. While for CSST, the spec-z error is several times larger than that measured from fibre spectroscopy, hence, we need to consider such effect. We show the damping parameter $\Sigma_z$ in the fifth column of Table \ref{tab:survey_params}.

The non-linear BAO signal in Eq. (\ref{eq:pkmu_bao}) is commonly modelled as \citep{Seo2007}
\begin{align}\label{eq:pbao_part}
    P_{\text{BAO, nl}}(k,\mu)=&\left[P_{\text{m, lin}}(k') - P_{\text{m, sm}}(k')\right]\times \\ \nonumber
    &\exp\left(-[k^2\mu^2\Sigma_{\|}^2 + k^2(1-\mu^2)\Sigma_{\perp}^2]/2\right), 
\end{align}
where $P_{\text{m, lin}}$ denotes the linear matter power spectrum, which is calculated from \textsc{camb}\footnote{https://camb.info/} \citep{Lewis2000}. $P_{\text{m, sm}}$ is the linear power spectrum without the BAO signal \citep{Eisenstein_Hu_1998}. $\Sigma_{\perp}$ and $\Sigma_{\|}$ are the pairwise rms Lagrangian displacements across and along the line of sight at the separation of the BAO scale \citep{Eisenstein_Seo_White_2007}. We estimate the displacements via 
\begin{align}
    \Sigma_{\perp}^2 &= 2\int \frac{dk}{6\pi^2}P_{\text{m, lin}}(k,\, z),\\
    \Sigma_{\|} &= (1 +f)\Sigma_{\perp}.
\end{align}

The non-linear BAO damping is mainly caused by the bulk flow and structure formation. Such effect not only smears the significance of the BAO signal in galaxy clustering, but also slightly shifts the BAO peak position, causing biased measurement on cosmological distances \cite[e.g.][]{Seo2005, Sherwin2012}. First proposed by \cite{Eisenstein2007b}, the density field reconstruction which inverses the Lagrangian displacement from the bulk flow, can approximately recovers the initial positions. As a result, it can largely reduce the BAO scale shifting from the non-linear evolution, and enhance the BAO scale detection significance. It has been widely studied in simulations and routinely used in real surveys \citep[e.g.][]{Padmanabhan2012, Seo2016, Beutler2017}. After reconstruction, the BAO damping parameters can be dramatically reduced. 
However, the CSST spec-z redshift error can be one order of magnitude larger than that from the fibre spectroscopy such as in DESI. It is not clear about the influence of redshift error on the reconstruction efficiency as a function of S/N \citep{White2010, Font-Ribera2014}. We expect that the spec-z systematic noise would also degrade the reconstruction efficiency. Therefore, we do not consider the improvement from the BAO reconstruction seriously in this study. We leave such investigation in future work. For the interest of the optimal case with reconstruction, we show the BAO constraints from the CSST spec-z survey with other surveys in Fig. \ref{fig:DA_Hz_diffsurveys}.

To measure galaxy clustering, e.g. power spectrum, from survey data, we need to assume a fiducial cosmology in order to convert angles and redshifts to distances. Due to the difference from the fiducial and true cosmologies, the observed    
angular and radial distances can be related to the real ones by the scale dilation parameters $\alpha_{\perp}$ and $\alpha_{\|}$ \citep{Anderson2014}, i.e.
\begin{align}
\alpha_{\perp} &= \frac{D_\text{A}(z) r_\text{d}^{\text{fid}}}{D_\text{A}^{\text{fid}}(z) r_\text{d}},\\
\alpha_{\|} &= \frac{H^{\text{fid}}(z) r_\text{d}^{\text{fid}}}{H(z) r_\text{d}},
\end{align}
where $\DA$ is the angular diameter distance and $H(z)$ is the Hubble parameter at redshift $z$, $\rd$ is the comoving sound horizon at the baryon-drag epoch $z_{\text{d}}$ with unity optical depth \citep{Hu_Sugiyama1996}, and the superscript ``fid'' denotes the fiducial cosmology. In Fourier space, the coordinates of power spectrum have the relation

\begin{align}
    k' &= \frac{k}{\alpha_{\perp}}\left[1+\mu^2\left(\alpha_{\perp}^2/\alpha_{\|}^2-1\right)\right]^{1/2},\\
    \mu' &= \mu \frac{\alpha_{\perp}}{\alpha_{\|}} \left[1+\mu^2\left(\alpha_{\perp}^2/\alpha_{\|}^2-1\right)\right]^{-1/2},
\end{align}
where $(k',\,\mu')$ and $(k, \mu)$ are the coordinates in the true and fiducial cosmologies, respectively.
From the constraints of $\alperp$ and $\alpara$, we can directly obtain the model independent constraints on $\DA/\rd$ and $H(z)\rd$, respectively\footnote{As the expected value of $\alperp$ and $\alpara$ equal to 1, we have $\sigma( \alperp)=\sigma(\ln (\DA/\rd))$ and $\sigma( \alpara)=\sigma(\ln (H(z)\rd))$. Throughout this paper, we use $\sigma(\DA/\rd)$ and $\sigma(H(z)\rd)$ to denote the fractional error of $\DA/\rd$ and $H(z)\rd$, respectively.}.

The scale dilation parameters affect the volume, the broad-band term from the Kaiser effect, as well as the isotropic power spectrum \citep{Samushia2011}. For the real data analysis of BAO, the scale dilation parameters in the volume and broad-band terms are highly degenerate with some nuisance parameters introduced to fit the broad-band power spectrum shape and any residual systematic noise; hence, they are weakly constrained. In other words, the scale dilation information is mainly constrained by the BAO signal. In our Fisher analysis without considering any nuisance parameters, we only include the scale dilation parameters in the linear power spectrum $P_{\text{m, lin}}$ and $P_\text{m, sm}$, i.e. the linear BAO power spectrum in Eq. (\ref{eq:pbao_part}), which is highlighted by the coordinate $k'$. We show the theoretical form of the derivative of $P_{\text{BAO, nl}}$ to the scale dilation parameters in Appendix \ref{sec:pbao_derivative_alphas}. 
Furthermore, we consider the effect of $P_{\text{sys}}$ as the systematic noise coming from the slitless spectroscopy with the sky background and star contamination \citep{Euclid2011}.

Overall, we set nine free parameters in the spec-z power spectrum for the BAO modelling, i.e. 
\begin{align}
    \theta = \{\alperp, \alpara, f, \Sigma_{\perp}, \Sigma_{\|}, \Sigma_{\text{FoG}}, \Sigma_z, b_g, P_{\text{sys}}\}.
\end{align}
For the photo-z power spectrum, we have the same set of parameters except for $P_{\text{sys}}$, i.e. ignoring the systematic noise of the photo-z data. When we forecast for the joint analyses, we distinguish the nuisance parameters, i.e. $\Sigma_{\perp}, \Sigma_{\|}, \Sigma_{\text{FoG}}, \Sigma_z, b_g$ for the spec-z and photo-z power spectra, respectively, even though some of them (e.g. $\Sigma_{\text{FoG}}$ and $b_g$) have the same value. For the modelling of the cross power spectrum, we describe it in Section \ref{sec:cross_pk}.
In the end, we obtain the one standard deviation ($1\sigma$) of $\alperp$ and $\alpara$ after marginalizing over the other parameters. We share our pipeline in \url{https://github.com/zdplayground/FisherBAO_CSST}. Our pipeline takes some reference from \textsc{GoFish}\footnote{https://github.com/ladosamushia/GoFish}.

\subsection{Numerical setting}
The Fisher forecast is sensitive to some numerical settings which we discuss in detail below.

\subsubsection{Galaxy bias}
For both the CSST spec-z and photo-z surveys, we adopt the linear galaxy bias as $b_g(z)=1+0.84z$ \citep{Weinberg2004}. We do not distinguish the galaxy biases between the spec-z and photo-z samples. Same as \cite{Lin2022}, we set a constant galaxy bias using the central redshift for each redshift bin. Considering the galaxy type from CSST spec-z would be ELGs dominated, we compare our bias with the galaxy bias of eBOSS ELGs \citep{Dawson2016} estimated from $b_g(z)=D(a)/D(0)$, which agrees within $10$ per cent.

\subsubsection{Systematic noise of spectroscopy}
Apart from the Poisson noise in the spec-z, there may be some systematic noise $\Psys$ originating from the slitless spectroscopic redshift measurement.  
For the fiducial and optimal analyses, we set $\Psys=0$. However, we take it as a free parameter and marginalize it over with other nuisance parameters when we report the constraints on $\DA/\rd$ and $H(z)\rd$, same as the process of \cite{Blanchard2020}. 

In addition, we study the influence from non-zero $\Psys$. As an additional noise term, with larger $\Psys$, the constraints on $\DA/\rd$ and $H(z)\rd$ from the spec-z clustering will be worse. Since we do not know the functional form of $\Psys$, as the simplest case, we assume it as scale and redshift independent, and vary it from $0$ to $10^4\Mpchcube$ as the optimistic to pessimistic cases. We can expect that the scale dependent $\Psys$ can smear the BAO scale with a more severe and complicated way than our case. 
Since our main goal is not to show the accurate Fisher forecast for the CSST spec-z BAO measurement, but to study the benefit from the joint analyses via cross-correlating the spec-z and photo-z clustering, such assumption is valid for our study.

\subsubsection{Imaging systematic effects}
In our study, we do not consider any influence on the galaxy power spectra from the imaging systematic effects. Since the sky coverage of CSST is very large, some foreground systematics such as the Galactic extinction, stellar density, and survey depth can induce spurious density fluctuation at large scales. We expect that applying imaging weight either from the linear multivariate regression \citep[e.g.][]{Ross2011,Ross2020} or the machine-learning algorithms \citep[e.g.][]{Rezaie2020, Razaie2023, Chaussidon2022} can effectively remove most of the imaging systematics at the BAO scale.

\subsubsection{Cross power spectrum}\label{sec:cross_pk}
In Section \ref{sec:bao_model}, we have shown the modelling of the galaxy auto power spectrum. In terms of the cross power spectrum between the spec-z and photo-z samples, we consider the differences of the redshift errors of the two density fields, i.e.
\begin{align}\label{eq:Pcross_bao}
    P_{g\text{, cross}} = & D^2(a) F_{\text{RSD, s}} F_{\text{RSD, p}} F_{\text{zerr, s}} F_{\text{zerr, p}}\times \nonumber\\
    & \left[P_{\text{BAO, nl}}(k,\mu) + P_{\text{m, sm}}(k)\right],
\end{align}
with
\begin{align}
    &P_{\text{BAO, nl}}(k,\mu)=\left[P_{\text{m, lin}}(k') - P_{\text{m, sm}}(k')\right]\times \\ \nonumber
    &\exp\left(-\big[k^2\mu^2(\Sigma_{\|\text{, s}}^2 + \Sigma_{\|\text{, p}}^2) + k^2(1-\mu^2)(\Sigma_{\perp\text{, s}}^2+\Sigma^2_{\perp\text{, p}})\big]/4\right), 
\end{align}
where the sub-scripts s and p denote spec-z and photo-z, respectively. We assume that there is no cross-correlation between the photo-z data and spectroscopic systematic noise, then the cross power spectrum does not have the influence from $\Psys$.
Overall, we have 14 free parameters in the joint analyses of the spec-z, photo-z and cross power spectra, i.e.
\begin{align}
    \theta = \{\alperp, \alpara, f, [\Sigma_{\perp}, \Sigma_{\|}, \Sigma_{\text{FoG}}, \Sigma_z, b_g]_x, \Psys\},
\end{align}
where $x=\{\text{s}, \text{p}\}$.

\subsubsection{Numerical derivative of the power spectrum}
In Fisher analysis, we need to do derivative of the power spectrum to the parameters. We can realize it either numerically or theoretically. We show the theoretical derivative of the logarithmic power spectrum to the parameters, as well as the inverse covariance matrix in Appendix \ref{sec:dlnP_dparams}. In terms of the numerical derivative, we increase and decrease each parameter by 1 per cent and use the symmetric difference, i.e.
\begin{align}
    P^{\prime}(x) = \frac{P(x+h) - P(x-h)}{2h},
\end{align}
where $x$ denotes the parameters in the power spectrum, and $h$ is the change of $x$, i.e. 1 per cent of $x$.
We have tested that such $h$ gives converged derivative.
In addition, we have checked that the Fisher forecast from the numerical and theoretical derivative are consistent to each other.

\subsubsection{Integration limit of Fisher matrix}
The upper limit of $k$ in the integration of Eq. (\ref{eq:fisher_single_tracer}) affects the final Fisher forecast significantly. \cite{Foroozan2021} has studied in detail on $k_{\text{max}}$ by comparing the Fisher-based predictions for the BAO measurements with the observational constraints from multiple spectroscopic surveys. For the anisotropic BAO scale measurement, we set $k_{\text{max}}=0.3\Mpch$, at which the result has converged well. We set $k_{\text{min}} = 2\pi/V^{1/3}_{\text{survey}}$, which is limited by the survey volume. Given the same sky coverage and redshift bin width, the survey volume is larger at higher redshift bin, hence, $k_{\text{min}}$ is smaller. 

For the $k$ integration, we set the step size $0.005\hMpc$. For the angular integration, we set $\mu$ from $-1.0$ to $1.0$ with step size $0.01$. We use the trapzoidal rule for the integration. We have tested different $k$ and $\mu$ step sizes, and confirmed that such setting makes the integration converge well.  

\section{Results}\label{sec:result}
\begin{table*}
	\centering
	\caption{Fisher forecast on the $1\sigma$ fractional errors of $\DA/\rd$ and $H(z) \rd$ (in per cent) constrained by the BAO measurements from the CSST galaxy surveys. As the fiducial case, we do not consider any systematic noise in the spec-z galaxy power spectrum. For each redshift bin, we consider the constraints from the spec-z and the photo-z auto power spectrum, respectively. In addition, we show the results from the spec-z power spectrum combined with the cross power spectra between the spec-z and photo-z data, as well as the joint analyses of the spec-z, photo-z auto and cross power spectra.}\label{tab:bao_forecast}
	\begin{tabular}{c c c c c |c| c c c c}
		\hline
		Redshift & \multicolumn{4}{|c|}{$\sigma (\DA/\rd)$ (per cent)} & & \multicolumn{4}{|c|}{$\sigma (H(z) \rd)$ (per cent)} \\ \cline{2-5} \cline{7-10}
		& spec-z & photo-z & spec-z+cross & spec-z+cross+photo-z & &spec-z & photo-z & spec-z+cross & spec-z+cross+photo-z \\
		\hline
0.0 < z < 0.2& 4.74 & 6.58 & 4.71 & 4.71 & & 12.21 & 90.48 & 12.20 & 12.20  \\
0.2 < z < 0.4& 1.74 & 2.55 & 1.72 & 1.72 & & 4.65 & 46.04 & 4.64 & 4.64  \\
0.4 < z < 0.6& 1.12 & 1.64 & 1.08 & 1.08 & & 3.01 & 37.35 & 2.99 & 2.99  \\
0.6 < z < 0.8& 0.84 & 1.22 & 0.80 & 0.80 & & 2.24 & 30.84 & 2.22 & 2.22  \\
0.8 < z < 1.0& 0.77 & 1.00 & 0.70 & 0.69 & & 1.99 & 27.55 & 1.96 & 1.96  \\
1.0 < z < 1.2& 1.41 & 0.90 & 0.93 & 0.77 & & 3.14 & 27.38 & 3.01 & 2.97  \\
1.2 < z < 1.4& 5.05 & 0.87 & 1.78 & 0.81 & & 9.45 & 29.31 & 8.65 & 8.37  \\
1.4 < z < 1.6& 24.52 & 0.83 & 3.92 & 0.81 & & 42.69 & 28.87 & 33.61 & 22.88  \\
		
        \hline
	\end{tabular}
\end{table*}

We show the Fisher forecast for the $1\sigma$ errors of $\DA/\rd$ and $H(z)\rd$ from the BAO measurements based on the CSST spec-z and photo-z galaxy clustering. For simplicity, we use the spec-z+cross to denote the joint analyses of the spec-z galaxy power spectrum and the cross power spectrum between the spec-z and photo-z samples. 
The spec-z+cross+photo-z denotes the combination of the spec-z+cross and the photo-z power spectrum. At last, we study the influence on the constraints of $\DA/\rd$ and $H(z)\rd$ from the spec-z systematic noise $\Psys$, the spec-z redshift success rate, and the photo-z error.

\subsection{Constraints on $\DA/\rd$ and $H(z)\rd$}

Table \ref{tab:bao_forecast} summarizes the $1\sigma$ constraints on $\DA/\rd$ and $H(z)\rd$ from different galaxy clustering tracers, i.e. the spec-z, photo-z, spec-z+cross, and spec-z+cross+photo-z at each redshift bin, respectively. As the fiducial and optimistic case, we do not assume any systematic noise in neither the spec-z nor the photo-z power spectra.

\begin{figure*}
    \centering
    \includegraphics[width=0.98\linewidth]{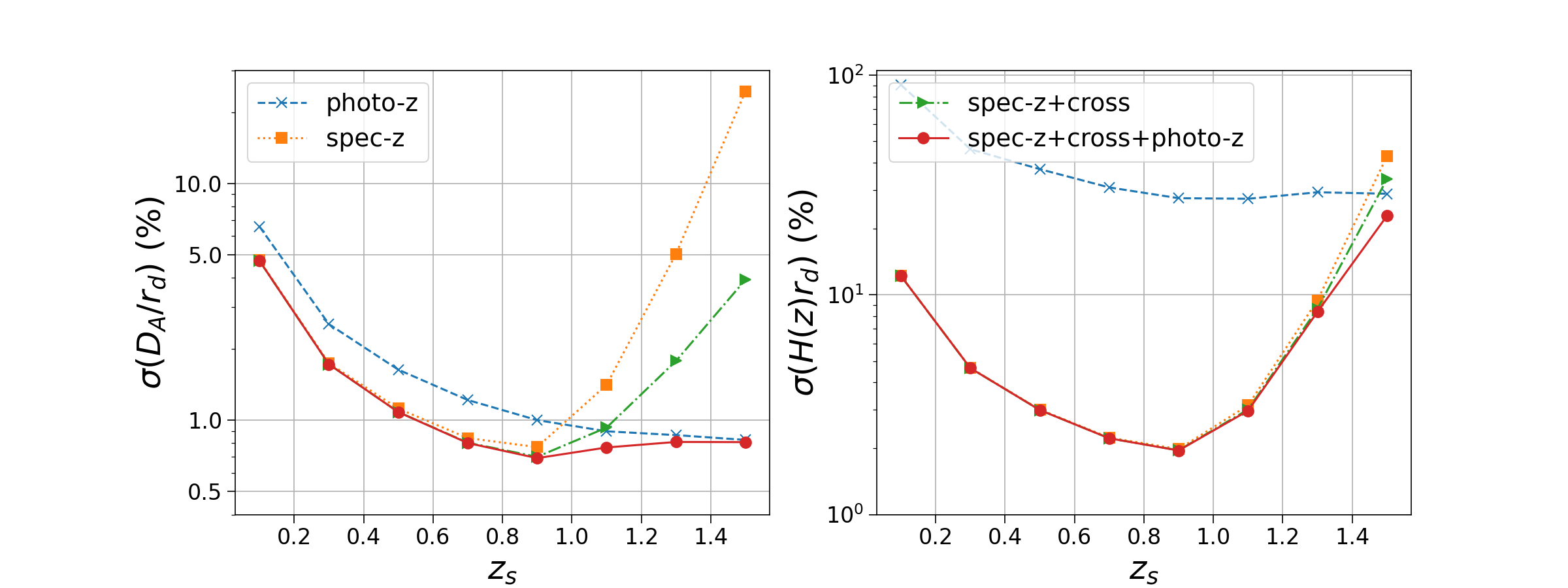}
    \caption{Fisher forecast of the $1\sigma$ constraints on $D_A(z)/\rd$ (in left panel) and $H(z)\rd$ (in right panel) from the BAO measurements based on the CSST spec-z and photo-z galaxy power spectra, as well as their joint analyses. Each type of markers represents one case, denoted in the legend. We do not take account of any systematic noise in the power spectra. For $D_A(z)/r_d$, the spec-z constraint increases as redshift increases until $z\simeq 1.0$, beyond which the shot noise dominates. For the photo-z constraint, it keeps increasing as $z$ is larger, and surpasses the spec-z one when $z>1.0$. At high redshifts, adding the cross-correlation between the spec-z and photo-z data can tighten the constraints from the spec-z alone, shown as the triangular and circular points. For $H(z)\rd$, due to large photo-z error, the dominant constraint is from the spec-z. The highest constraint is about $2$ per cent at $0.8<z<1.0$.}
    \label{fig:sigma_DA_Hz_default}
\end{figure*}

Fig. \ref{fig:sigma_DA_Hz_default} shows the constraints on $\DA/\rd$ and $H(z)\rd$ at different redshift bins from Table \ref{tab:bao_forecast}. The left panel is for $\DA/\rd$. Different markers depict cases from a single tracer or the joint analyses. 
For the spec-z tracer, shown as the squares, at $z<1$, the $1\sigma$ error of $\DA/\rd$ decreases as redshift increases, thanks to the increase of survey volume and the relatively high galaxy number density. It can reach sub-per cent level at $0.6<z<1.0$ without the BAO reconstruction.
We expect that the BAO reconstruction would further improve the constraints, especially at lower redshifts with higher galaxy number densities and smaller redshift error.
At $z>1$, since the spec-z galaxy number density decreases dramatically, hence, the shot noise begins to dominate, and the constraint decreases quickly as redshift goes larger. 

In terms of the constraint from the photo-z tracer, it keeps increasing as redshift increases (until $z\sim 2.0$ from Fig. \ref{fig:photoz_sigma_DA}), shown as the crosses in Fig. \ref{fig:sigma_DA_Hz_default}, which is again due to the increasing survey volume and relatively high galaxy number density even at high redshifts. 
When $z>1$, the photo-z constraint on $\DA/\rd$ becomes better than that from the spec-z.
Of course, such conclusion highly depends on our assumptions about the redshift errors and systematic noises of the surveys. In addition, we have ignored the redshift outlier fraction of the photo-z survey, which can also reduce the S/N. As \cite{Liu2023} showed that the CSST photo-z outlier fraction is about 8 per cent, we have tested such effect on the photo-z and cross power spectra, and found that the change on our fiducial forecast is negligible, which is consistent to the conclusion of \cite{Ansari2019}. 

For the constraints on $\DA/\rd$ at high redshifts ($z>1$), there is significant gain from the cross-correlation between the spec-z and photo-z data compared to that from the spec-z alone, shown as the triangles in Fig. \ref{fig:sigma_DA_Hz_default}. The improvement can be calculated from $1-\sigma_{\text{spec-z+cross}}/\sigma_{\text{spec-z}}$. It is larger than 30 per cent at $1.0<z<1.2$ for the case spec-z+cross, and even larger with the inclusion of the photo-z clustering. At larger redshifts, the increased constraints can be a few times tighter from the joint analyses compared to the spec-z one which decreases quickly due to the drop-off in the number density. Such improvement should be valid even if there are systematic noises in the spec-z and photo-z data. Because the noise from one data set is unlikely to correlate with the signal or noise of the other data set, the noises from the two data sets contaminate the cross-correlation signal little. Without any systematic noise in the photo-z galaxy clustering, the constraint on $\DA/\rd$ is significantly higher than the spec-z one at $z>1$, hence, the constraint of spec-z+cross+photo-z is dominated by the photo-z one, shown as the circular points. 

The right panel of Fig. \ref{fig:sigma_DA_Hz_default} shows the constraints on $H(z)\rd$. The spec-z constraint (as a function of redshift) has a similar shape as that of $\DA/\rd$. The redshift accuracy is vital for the constraint on $H(z)\rd$ which traces information along the line of sight. Due to large photo-z error, there is little constraint from the photo-z tracer, as expected. Therefore, the constraint is dominated by the spec-z one in the cases of joint analyses. 

\subsection{Dependence on the spec-z systematic noise}

\begin{figure*}
    \centering
    \includegraphics[width=0.9\linewidth]{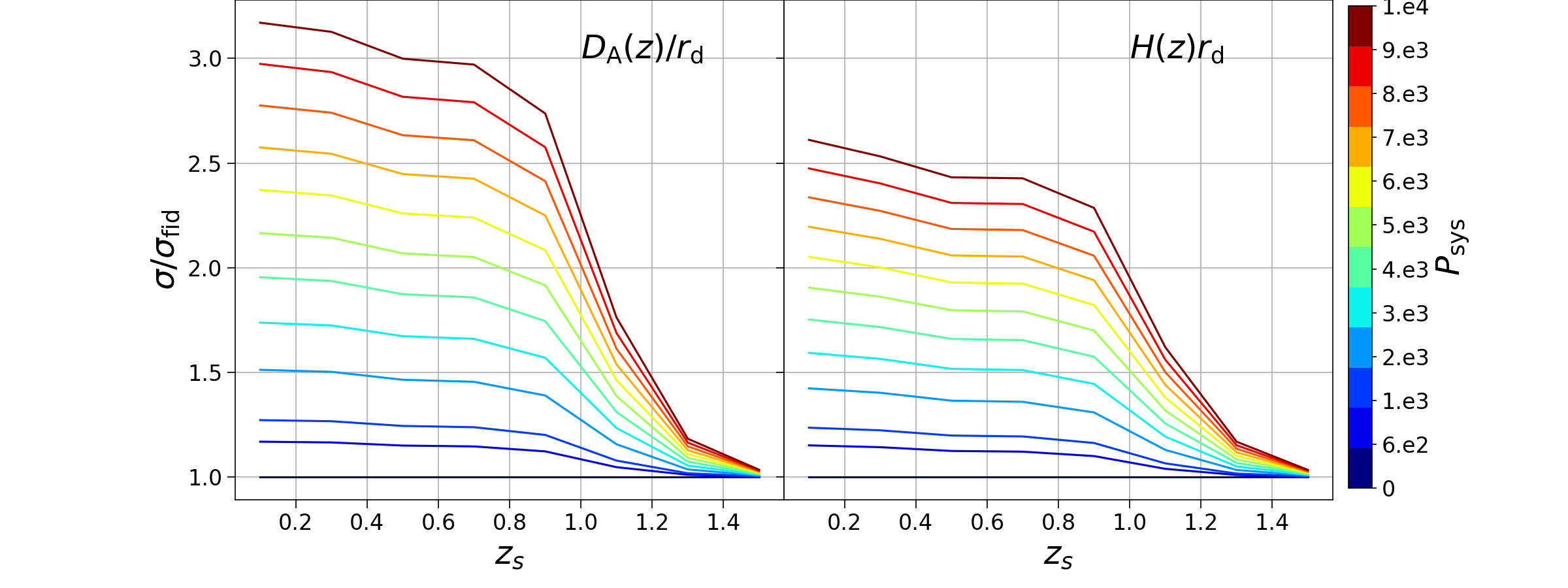}
    \caption{Reduction of the constraints on $\DA/\rd$ and $H(z)\rd$ due to the systematic noise $\Psys$ in the spec-z galaxy power spectrum. We show the ratio of the $1\sigma$ constraints from the spec-z tracer with and without considering $\Psys$. The left panel is for $\DA/\rd$, and the right panel is for $H(z)\rd$. Different colours denote different values of $\Psys$ considered. We vary the systematic noise from $0$ to $10^4\Mpchcube$. With larger $\Psys$, $\sigma/\sigma_{\text{fid}}$ becomes larger.}
    \label{fig:sigma_DA_Hz_diffPsys}
\end{figure*}

\begin{figure}
    \centering
    \includegraphics[width=0.9\linewidth]{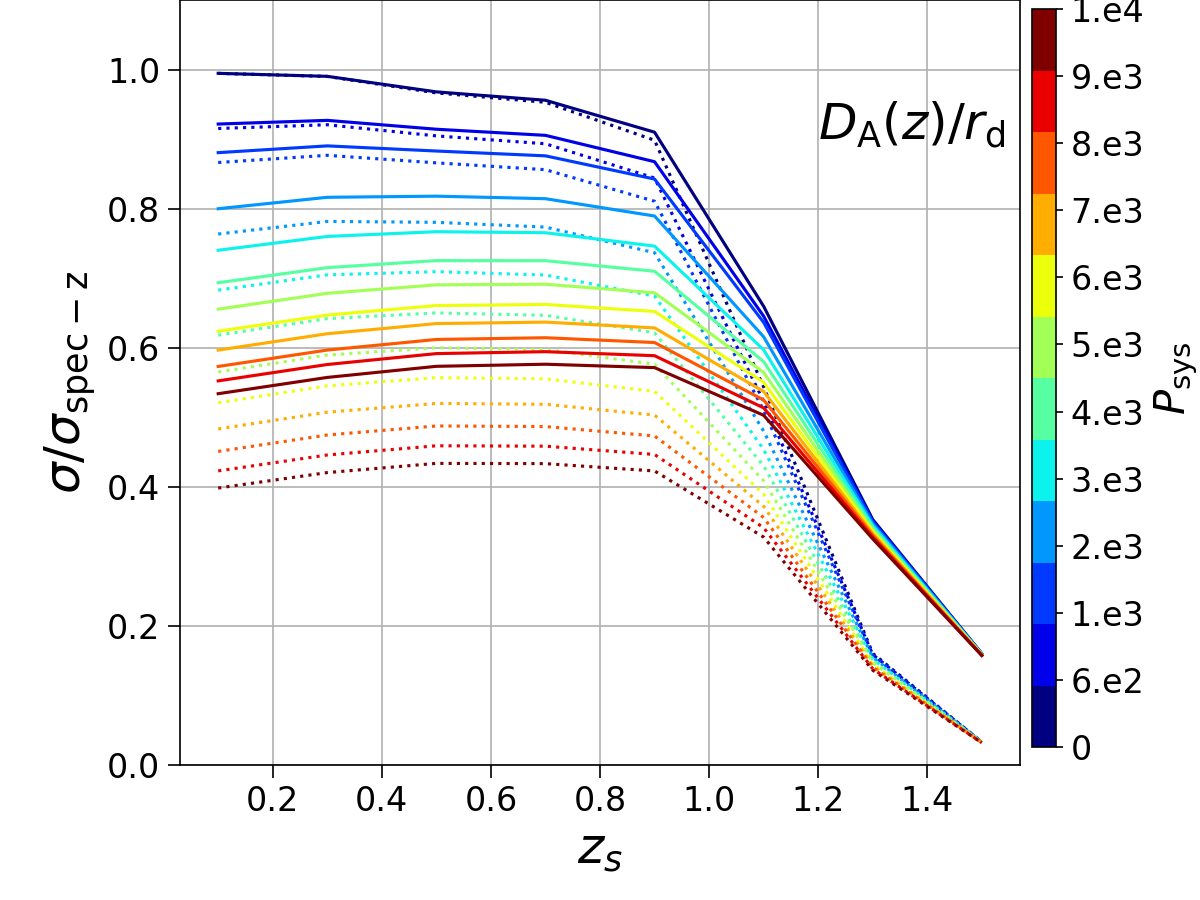}
    \caption{Increase of the constraints on $D_A(z)/\rd$ from the spec-z+cross compared to that from the spec-z tracer, shown as the solid lines. Different colours indicate different systematic noises $\Psys$ considered in the spec-z power spectrum. With larger $\Psys$, $\sigma/\sigma_{\text{spec-z}}$ becomes smaller. We overplot the results from the spec-z+cross+photo-z, shown as the dotted lines.}
    \label{fig:sigma_DA_specz_cross}
\end{figure}

The exact constraints on $\DA/\rd$ and $H(z)\rd$ from the CSST spec-z and photo-z galaxy clustering will highly depend on the systematic noises. $\Psys$ propagates to the power spectrum covariance matrix, and affects the constraints on $\DA/\rd$ and $H(z)\rd$.

Currently, we have little knowledge on the systematic noise.
For the simplest case, we consider the spec-z systematic noise $\Psys$ as a constant, and vary it from $0$ to $10^4\Mpchcube$.
In Fig. \ref{fig:sigma_DA_Hz_diffPsys}, we show the damping effect on the constraints of $\DA/\rd$ and $H(z)\rd$ from the spec-z tracer with $\Psys$. Different colours represent different values of $\Psys$. As expected, with larger $\Psys$, the BAO S/N becomes lower, hence, the constraint becomes weaker compared to that of the fiducial case with $\Psys=0$.
Interestingly, the systematic reduction affects $\DA/\rd$ larger than $H(z)\rd$. This is because that $H(z)$ is sensitive to the modes along the line of sight. Due to the Kaiser effect, the power spectrum signal is larger from the line-of-sight modes, hence, the S/N is larger than that from the modes perpendicular to the line of sight which $\DA$ is sensitive to. 
In addition, at $z>1$, as the spec-z shot noise increases quickly, the influence from the systematic noise becomes relatively mild.

As noted before, the cross-correlation between the spec-z and photo-z clustering is likely to remove much of the systematics which usually do not correlate with each other from the spec-z and photo-z data sets. Therefore, it is meaningful to see how much improvement of the constraints on $\DA/\rd$ from the addition of the cross-correlation with respect to that from the spec-z tracer alone.
Fig. \ref{fig:sigma_DA_specz_cross} shows the comparison of the $1\sigma$ constraint on $\DA/\rd$ from the spec-z+cross and spec-z with different systematic noises. Given some $\Psys$, the cross-correlation can effectively improve the constraints over all the redshift range. It gives larger gain at higher redshift, which is simply due to the low constraints from the spec-z tracer. For example, Fig. \ref{fig:sigalperp_specz_cross_Psys2.e3} shows the constraints on $\DA/\rd$ from the case with $\Psys=2\times 10^3\Mpchcube$.

With larger $\Psys$, the improvement is also more significant. From Fig. \ref{fig:sigma_DA_specz_cross}, we see that if $\Psys>2\times 10^3 \Mpchcube$, the improvement from the joint analyses spec-z+cross is larger than $\sim 20$ per cent at all redshifts; if $\Psys>8\times 10^3 \Mpchcube$, it is larger than 40 per cent. For comparison, we also show the results from the spec-z+cross+photo-z as the dotted lines. Adding the photo-z clustering can further increase the constraint compared to that of the spec-z+cross, especially for the case with larger spec-z $\Psys$. However, this also highly depends on the photo-z systematic noise which is not considered in our study.

For the constraints on $H(z)\rd$, adding the cross-correlation barely improves the constraint from the spec-z clustering. We have checked that the gain is less than 10 per cent from the spec-z+cross joint analyses at $z<1.0$, even for the case with a large $\Psys$, e.g. $10^4\hMpccube$, in the spec-z power spectrum.

\subsection{Dependence on the spectroscopic redshift success rate}
\begin{figure}
    \centering
    \includegraphics[width=0.98\linewidth]{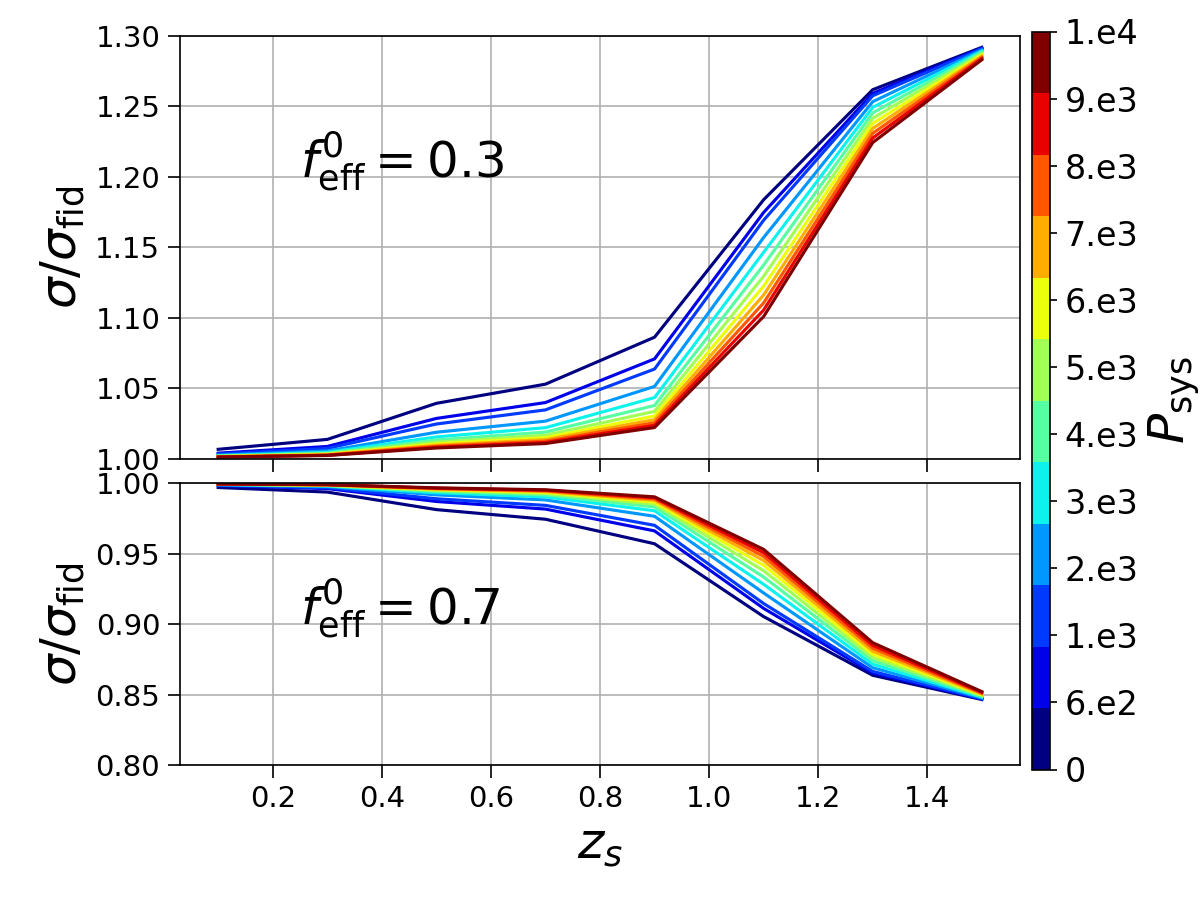}
    \caption{Dependence of the $\DA/\rd$ constraints on the redshift success rate. We divide the constraint from the spec-z+cross joint analyses with a lower or higher redshift success rate in the spec-z data over that with the fiducial one. The upper (lower) panel is for the case with a lower (higher) redshift success rate. Different colours denote different $\Psys$ considered in the spec-z power spectrum. The effect of $\Psys$ on $\sigma/\sigma_{\text{fid}}$ is small; with smaller $\Psys$, $\sigma/\sigma_{\text{fid}}$ deviates from 1 slightly larger.} \label{fig:sigma_DA_specz_cross_diff_f0eff}
\end{figure}

Given some level of the redshift error, varying the spec-z success rate changes the galaxy number density, then affects the BAO constraints systematically. To check such effect, we decrease and increase the redshift success rate $f_{\text{eff}}^z=f^0_{\text{eff}}/(1+\zs)$ by 40 per cent, i.e. via replacing $f^0_{\text{eff}}=0.5$ by $0.3$ and $0.7$, respectively. Fig. \ref{fig:sigma_diff_f0eff} shows the change of the constraints on $\DA/\rd$ and $H(z)\rd$ from the spec-z power spectrum with different redshift success rates. 

We are particularly interested in the influence of the redshift success rate on the $\DA/\rd$ constraints from the joint analyses spec-z+cross. Fig. \ref{fig:sigma_DA_specz_cross_diff_f0eff} shows the ratio of the spec-z+cross constraints with a lower or higher redshift success rate in the spec-z data compared to the fiducial one. If the ratio deviates more from unity, the influence is larger. It increases as redshift becomes larger. At a given redshift, with larger $\Psys$, the influence is smaller. Overall, the influence on the constraints from different redshift success rates is not very significant, within 30 per cent and 15 per cent for the cases with the lower and higher redshift success rates, respectively.

\subsection{Dependence on the photo-z error}
\begin{figure}
    \centering
    \includegraphics[width=0.9\linewidth]{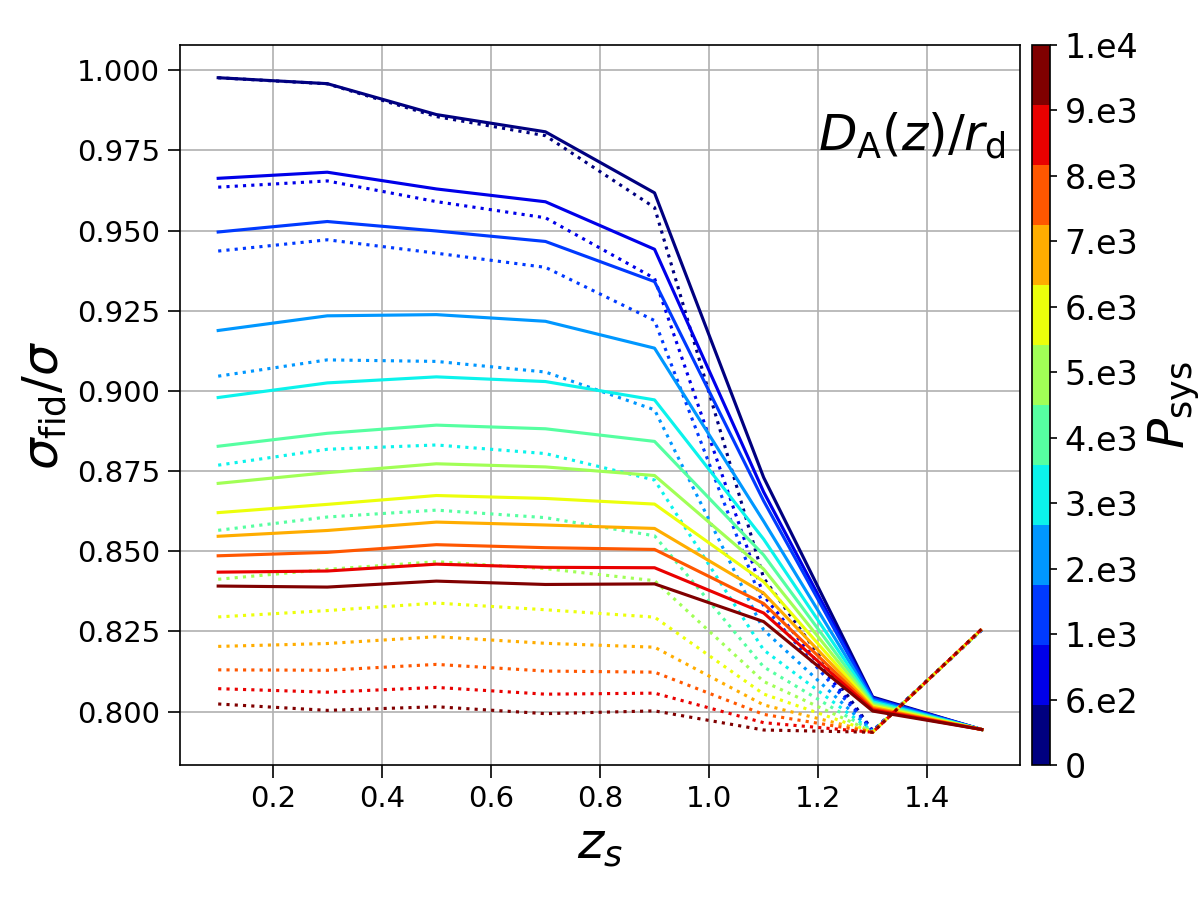}
    \caption{Dependence of the $\DA/\rd$ constraints on the redshift error of the photo-z sample in the joint analyses. We show the ratio of the constraints from the fiducial case over that with larger photo-z error $\sigma_z=0.05(1+z)$. The solid lines are the results of the spec-z+cross, and the dotted lines are from the spec-z+cross+photo-z. Increasing the photo-z error reduces the constraints from the joint analyses. With the existence of $\Psys$, $\sigma_{\text{fid}}/\sigma$ becomes smaller as $\Psys$ is larger.}
    \label{fig:sigma_DA_photozerr}
\end{figure}

Furthermore, we study the influence of the photo-z error on the constraints of $\DA/\rd$ from the joint analyses. We replace the fiducial photo-z sample by the one with larger photo-z error $\sigma_z=0.05(1+z)$. As we have compared the two photo-z samples in Fig. \ref{fig:nz_phot}, the number density is about $40$ -- $60$ per cent lower for the fiducial one with smaller photo-z error. Based on these two photo-z samples, we show the constraints on $\DA/\rd$ from the photo-z galaxy clustering extending to $z=4$ in Fig. \ref{fig:photoz_sigma_DA}.    

In Fig. \ref{fig:sigma_DA_photozerr}, the solid lines show the ratio of the constraints on $\DA/\rd$ from the spec-z+cross analyses with smaller redshift error in the photo-z sample over that with larger photo-z error. Even though the galaxy number density is lower for the sample with smaller photo-z error, the shot noise is still less significant compared to the cosmic variance. Therefore, using the sample with smaller photo-z error gives better performance for the joint analyses. If we include the photo-z clustering to have the spec-z+cross+photo-z, the best constraint can be further improved, shown as the dotted lines. However, the improvement is not very significant, $\leq 20$ per cent for the $\Psys$ range that we consider. The benefit on constraining $\DA/\rd$ from the joint analyses is robust even if we do not have the photo-z sample with the smaller redshift error.

\subsection{Comparison with other Stage-IV spectroscopic surveys}
\begin{figure*}
    \centering
    \includegraphics[width=0.99\linewidth]{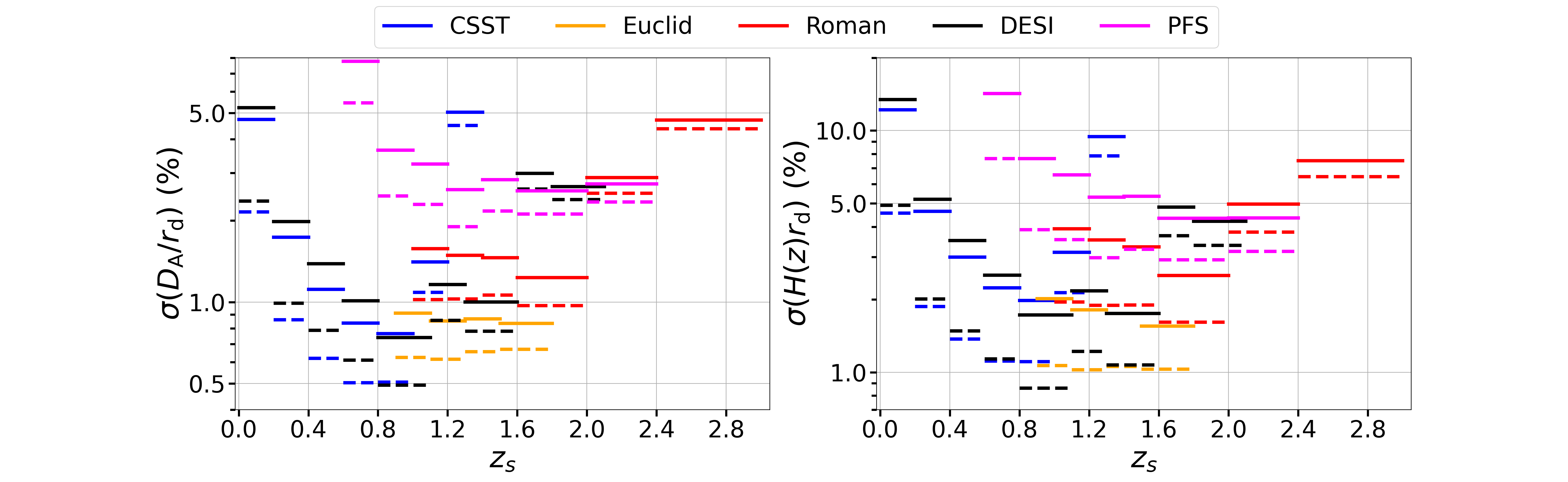}
    \caption{Comparison of $1\sigma$ constraints on $\DA/\rd$ and $H(z)\rd$ via the BAO measurements from different Stage-IV spectroscopic surveys. The solid lines denote the results before the BAO reconstruction, and the dashed lines indicate the cases after reconstruction. For DESI, we do not include the constraints from Ly $\alpha$ forest at $z>2.1$.}
    \label{fig:DA_Hz_diffsurveys}
\end{figure*}
Based on our Fisher forecast pipeline\footnote{Here, we set $k_{\text{max}}=0.5\hMpc$, which is commonly chosen in other literatures for the BAO forecast.}, we calculate and compare the BAO constraints from the current and forthcoming Stage-IV spectroscopic redshift surveys, including CSST, Euclid, Roman, DESI, and PFS. The former three are space telescopes with slitless spectrograph, and the latter two are ground based telescopes with fibre-fed spectrograph.

For the Euclid spectroscopic survey, it is able to detect larger than 30 million H$\alpha$ emitters in the redshift range $0.9<z<1.8$ with the sky coverage 15000 deg$^2$ \citep{Pozzetti2016}. The redshift error is $\sim 0.001(1+z)$. The expected number density and galaxy bias can be found in \cite{Blanchard2020}, and the redshift bins are set as [0.9, 1.1], [1.1, 1.3], [1.3, 1.5], and [1.5, 1.8].

For the Roman High Latitude Spectroscopic Survey, it can map 2000 deg$^2$ sky area and measure $\sim 10$ million ELG redshifts with H$\alpha$ emission lines in $1<z<2$ and $\sim 2$ million with [O III] emission lines in $2<z<3$ \citep{Wang2022}. The redshift error is $\sim 0.001(1+z)$. We take the galaxy number distributions  of H$\alpha$ and [O III] from table 1 and 2 of \cite{Wang2022}, respectively. The distributions are based on the dust model with $A_V=1.92$ and the flux limit $1\times 10^{-16}$ erg s$^{-1}$cm$^{-2}$. The galaxy bias model is adopted from \cite{Zhai2021b}. We set the redshift bins as [1.0, 1.2], [1.2, 1.4], [1.4, 1.6], [1.6, 2.0], [2.0, 2.4], and [2.4, 3.0].

For DESI, it observes four types of galaxies, i.e., bright galaxy samples (BGS; $0<z<0.4$), LRGs ($0.4<z<1.1$), ELGs ($1.1<z<1.6$), and quasi-stellar objects (QSO; $z>1.6$) at different redshift ranges. QSO with $z>2.1$ are used for the study of Ly$\alpha$ forest, which are not considered in our forecast. We take reference of the galaxy number density and galaxy bias for each extragalactic tracer from the DESI One-Percent Survey \citep{Adame2023a, Adame2023b}. The redshift error of
each tracer is below $0.0005(1+z)$, and the sky coverage is 14000 deg$^2$. To have overlap with the redshift bins from other surveys, we set the DESI redshift bins as [0, 0.2], [0.2, 0.4], [0.4, 0.6], [0.6, 0.8], [0.8, 1.1], [1.1, 1.3], [1.3, 1.6], [1.6, 1.8], and [1.8, 2.1].

For the PFS spectroscopic survey, it observes ELGs with [O II] emission lines in the redshift range $0.8<z<2.4$ with the survey area over 1400 deg$^2$. The redshift error is below $0.0007(1+z)$. We take reference of the redshift bins, galaxy number density, and galaxy bias from table 2 of \cite{Takada2014}.

Fig. \ref{fig:DA_Hz_diffsurveys} compares the $1\sigma$ constraints on $\DA/\rd$ and $H(z)\rd$ from CSST, Euclid, Roman, DESI, and PFS. The solid lines denote the constraints from the BAO measurements before the density field reconstruction. Different colours represent different surveys. The line length gives the redshift bin size. Since the BAO reconstruction has been routinely adopted in previous spectroscopic surveys, we also forecast the BAO constraints after reconstruction shown as the dashed lines, though we have not considered any systematic errors from the slitless spectroscopy of CSST, Euclid and Roman. We only consider the influence of shot noise on reconstruction; we apply a reduction scaling factor to the BAO signal damping parameters $\Sigma_{\perp}$ and $\Sigma_{\parallel}$ \citep{White2010, Font-Ribera2014}. In addition, we set the Fingers-of-God damping parameter $\Sigma_{\text{FoG}}$ close to 0 after reconstruction.  Therefore, our forecast should be taken as an optimistic case. To check the pipeline accuracy, we have compared our DESI forecast with that of \cite{Adame2023a} setting the same redshift bin size $0.1$ for each tracer. The relative differences are mostly within 3 per cent for both $\DA/\rd$ and $H(z)\rd$ constraints from BGS, LRG and ELG\footnote{While for QSO, the relative difference is $\sim 9$ per cent, which is a little bit large and may require further investigation.}. The discrepancy can be due to some differences from the fiducial cosmologies, the model parameters and the forecast settings in the two cases. Overall, our forecast is generally reliable, and gives an overview on the BAO constraints from the ongoing and future Stage-IV spectroscopic surveys. For CSST, it has potential to give tighter constraints on the BAO scale than that from DESI at $z<0.8$, thanks to its larger sky coverage and higher galaxy number density.

\section{Conclusions}\label{sec:conclusions}
As one of the Stage IV galaxy surveys, CSST will perform the photo-z imaging survey and slitless spec-z survey simultaneously. The two surveys will cover the same fraction of sky area (17\ 500 $\text{deg}^2$), and the maximum redshift can reach $4.0$ and $1.5$ from the CSST photo-z and spec-z surveys, respectively. In this study, we provide a Fisher forecast on the constraints of $\DA/\rd$ and $H(z)\rd$ based on the BAO scale measurements, focusing on the improvement from cross-correlating the photo-z and spec-z samples over that from the spec-z alone.

We first model the galaxy redshift distribution for the CSST surveys. For the photo-z sample, we adopt the mock from \cite{Cao2018}, which is constructed from the COSMOS catalogue \citep{Capak2007, Ilbert2009}. For the spec-z distribution, we construct it based on the zCOSMOS catalogue \citep{Lilly2007, Lilly2009}. We consider the redshift range $0<z<1.6$, beyond which the zCOSMOS sample is too sparse to model the distribution. We divide the redshift range into eight uniform bins, and does the same for the photo-z sample. Based on the mock galaxy redshift distributions, we can estimate the galaxy shot noise at each redshift bin for both surveys.
 Then we construct the anisotropic galaxy power spectrum, taking account of the RSD, galaxy bias, BAO damping scales, redshift error (for both the spec-z and photo-z samples), as well as the systematic noise from the slitless spectroscopy. 
We model the cross power spectrum taking account of different redshift errors of the two surveys. For the BAO constraints, we only focus on relatively large scales with $k<0.3 \hMpc$.

For the fiducial case without including any systematic noise in the spec-z galaxy power spectrum, the BAO scale measurement can constrain $\DA/\rd$ in sub-per cent level at $0.6<z<1.0$ before the BAO reconstruction. At $z>1.0$, as the spectroscopic galaxy sample becomes sparse, the spec-z constraint decreases quickly. 
For CSST, the constraint from photo-z increases with increasing redshift until $z\sim 2$, and surpasses the spec-z constraint at $z\sim 1.0$ in the case without any systematic noises.
Cross-correlating the spec-z sample with the much denser photo-z sample can significantly improve the constraints on $\DA/\rd$ at $z>1$.

As a main goal, we quantify the increase on the constraints of $\DA/\rd$ from the joint analyses of the spec-z and photo-z clustering. 
The main result is shown in Fig \ref{fig:sigma_DA_Hz_default}.
The improvement is larger than 30 per cent at $1.0<z<1.2$, and even larger at higher redshifts. It is because that the constraint from the photo-z clustering starts to dominate at $z>1$, which however depends on the quality removing the imaging systematics.
We also check the constraints on $H(z)\rd$ from the spec-z clustering and the joint analyses. 
The improvement is very mild as expected.

We consider different systematic effects on the improvement of the joint analyses, including the spec-z systematic noise, the spec-z redshift success rate, and the photo-z error.
With larger systematic noise in the spec-z data, the improvement on the $\DA/\rd$ constraint from the cross-correlation is more significant, since the systematic noise suppresses the S/N of the spec-z data but does not influence the cross-correlation.
The influence from varying the redshift success rate is not significant, e.g. varying within 30 per cent and 15 per cent if we lower or increase the fiducial redshift success rate by 40 per cent, respectively.
Using the photo-z sample with smaller redshift error gives slightly better constraints on $\DA/\rd$.
Overall, the cross-correlation between the spec-z and photo-z clustering improves the BAO constraint from the spec-z alone, especially at higher redshifts. The improvement of the joint analyses is robust from the systematics that we have considered.

For the comparison with the BAO constraints from CSST, we apply our pipeline forecasting the BAO constraints from other Stage-IV spectroscopic surveys, including Euclid, Roman, DESI and PFS. It gives an overview on the BAO constraints from these surveys. Specifically, with the larger survey area and higher galaxy number density, CSST has potential to provide tighter constraints than DESI at $z<0.8$.
We expect that our study can be beneficial for the future CSST BAO analysis on real data. We can apply the study to other galaxy surveys which conduct both spec-z and photo-z surveys, such as Euclid and Roman, as well as a joint analysis of a spec-z survey and a photo-z survey which cover the same survey volume.

\section*{Acknowledgements}
We thank Ye Cao and Yan Gong for providing us the CSST photo-z galaxy distribution which is based on the COSMOS catalogue. We thank the referee for the constructive and detailed comments to greatly improve the manuscript.
ZD thank Yan Gong, Yulin Jiao, Hee-Jong Seo, Lado Samushia, Zhao Chen, Ji Yao, and Huanyuan Shan for helpful discussions. 
ZD and YY were supported by the National Key Basic Research and Development Program of China (grant number 2018YFA0404504) and the National Science Foundation of China (grant numbers 12273020, 11621303, and 11890691), and the China Manned Space Project with number CMS-CSST-2021-A03. This work made use of the Gravity Supercomputer at the Department of Astronomy, Shanghai Jiao Tong University.
%%%%%%%%%%%%%%%%%%%%%%%%%%%%%%%%%%%%%%%%%%%%%%%%%%
\section*{Data Availability}
We share our pipeline and input data in \url{https://github.com/zdplayground/FisherBAO_CSST}.

\bibliographystyle{mnras}
\bibliography{references} 

\appendix
%%%%%%%%%%%%%%%%% APPENDICES %%%%%%%%%%%%%%%%%%%%%

\appendix

\section{Logarithmic derivative of the BAO power spectrum to the scale dilation parameters}\label{sec:pbao_derivative_alphas}
From Eq. (\ref{eq:pbao_part}), we can obtain the logarithmic derivative of the BAO power spectrum with respective to the scale dilation parameters, assuming the fiducial cosmology is close to the true, i.e.
\begin{align}
    \frac{\partial \text{ln} P_{\text{BAO, nl}}}{\partial \alpara}\bigg\vert_{\alperp,\, \alpara=1} &= \frac{\partial \text{ln} P_{\text{BAO, nl}}}{\partial k'} \frac{\partial k'}{\partial \alpara}=-\mu^2 \frac{\partial \text{ln} P_{\text{BAO, nl}}}{\partial \text{ln} k'},\nonumber \\
    &= -\mu^2 \frac{\partial \text{ln} P_{\text{BAO, lin}}}{\partial \text{ln} k'}
\end{align}
and
\begin{align}
    \frac{\partial \text{ln} P_{\text{BAO, nl}}}{\partial \alperp}\bigg\vert_{\alperp,\, \alpara=1} =-(1-\mu^2)\frac{\partial \text{ln} P_{\text{BAO, lin}}}{\partial \text{ln} k'}.
\end{align}
Similar as in section 4.2 of \cite{Seo2007}, we can examine the spherical symmetry test for the Fisher matrix of $\alperp$ and $\alpara$ by assuming the isotropic BAO damping and ignoring the RSD effect and the redshift error. The cross-correlation coefficient of $\alperp$ and $\alpara$ would be $-0.41$.

\section{Theoretical derivative of the Fisher matrix}\label{sec:dlnP_dparams}
We rewrite the Fisher matrix for the multitracer case as 
\begin{align}\label{eq:Fish_lnP}
    \text{F}_{ij} = \int_{-1}^{1}\int_{k_{\text{min}}}^{k_{\text{max}}} \frac{2\pi k^2 dk d\mu}{2(2\pi)^3} V_{\text{survey}} \frac{\partial\,\text{ln} P^T(k,\mu)}{\partial \theta_i} \text{D}^{-1} \frac{\partial\, \text{ln} P(k,\mu)}{\partial \theta_j}, 
\end{align}
with 
\begin{align}\label{eq:inv_cov_lnP}
\text{D}^{-1}= \frac{1}{(\hPs \hPp - \Psp^2)^2}
\begin{bmatrix} 
\Ps^2\hPp^2 & -2\Ps \Psp^2 \hPp & \Ps P_{\text{sp}}^2 \Pp \\
-2\Ps \Psp^2 \hPp & 2(\hPs\hPp + \Psp^2)\Psp^2 & -2\hPs\Psp^2\Pp \\
\Ps \Psp^2 \Pp & -2\hPs \Psp^2 \Pp & \hPs^2 \Pp^2
\end{bmatrix},
\end{align}
where the hat sign denotes the shot noise included. Substituting Eq. (\ref{eq:inv_cov_lnP}) in Eq. (\ref{eq:Fish_lnP}) and calculating each term of the Fisher matrix should match to the results in the appendix of \cite{Zhao2016}.

In terms of the auto galaxy power spectrum given by Eq. (\ref{eq:pkmu_bao}) and without considering the systematic noise, we have the logarithmic derivative of the power spectrum to the parameters as
\begin{align}
    \frac{\partial \text{ln}P_g}{\partial \alperp} &=-(1-\mu^2)\frac{\partial\text{ln} (P_{\text{BAO, nl}} + P_{\text{sm}})}{\partial \text{ln}k'},\\
    \frac{\partial \text{ln}P_g}{\partial \alpara} &=-\mu^2\frac{\partial\text{ln} (P_{\text{BAO, nl}} + P_{\text{sm}})}{\partial \text{ln}k'},\\
    \frac{\partial \text{ln}P_g}{\partial \Sigma_{\text{FoG}}} &=\frac{4}{\Sigma_{\text{FoG}}}(F_{\text{FoG}}-1),\\
    \frac{\partial \text{ln}P_g}{\partial \Sigma_{z}} &=-2k^2\mu^2 \Sigma_z,\\
    \frac{\partial \text{ln}P_g}{\partial \Sigma_{\perp}} &= -k^2(1-\mu^2)\Sigma_{\perp} \frac{P_{\text{BAO, nl}}}{P_{\text{BAO, nl}} + P_{\text{sm}}},\\
    \frac{\partial \text{ln}P_g}{\partial \Sigma_{\parallel}} &= -k^2 \mu^2\Sigma_{\parallel} \frac{P_{\text{BAO, nl}}}{P_{\text{BAO, nl}} + P_{\text{sm}}},\\
    \frac{\partial \text{ln}P_g}{\partial f} &= \frac{2\mu^2}{b_g + f\mu^2}\\
    \frac{\partial \text{ln}P_g}{\partial b_g} &= \frac{2}{b_g + f\mu^2}.
\end{align}
For the case with the systematic noise considered in $P_g$, we just need to add a multiplication factor $P_g/(P_g + P_{\text{sys}})$ on the right of the above equations. In addition, we consider $\Psys$ as a free parameter even in the case $\Psys=0$, and have the derivative as
\begin{align}
   \frac{\partial \text{ln}(P_g+\Psys)}{\partial \Psys} &= \frac{1}{P_g + \Psys}. 
\end{align}

Similarly, we can derive the logarithmic derivative of the cross power spectrum.
%%%%%%%%%%%%%%%%%%%%%%%%%%%%%%%%%%%%%%%%%%%%%%%%%%

\section{Constraints on $\DA/\rd$ from the photo-z survey}
\begin{figure}
    \centering
    \includegraphics[width=0.9\linewidth]{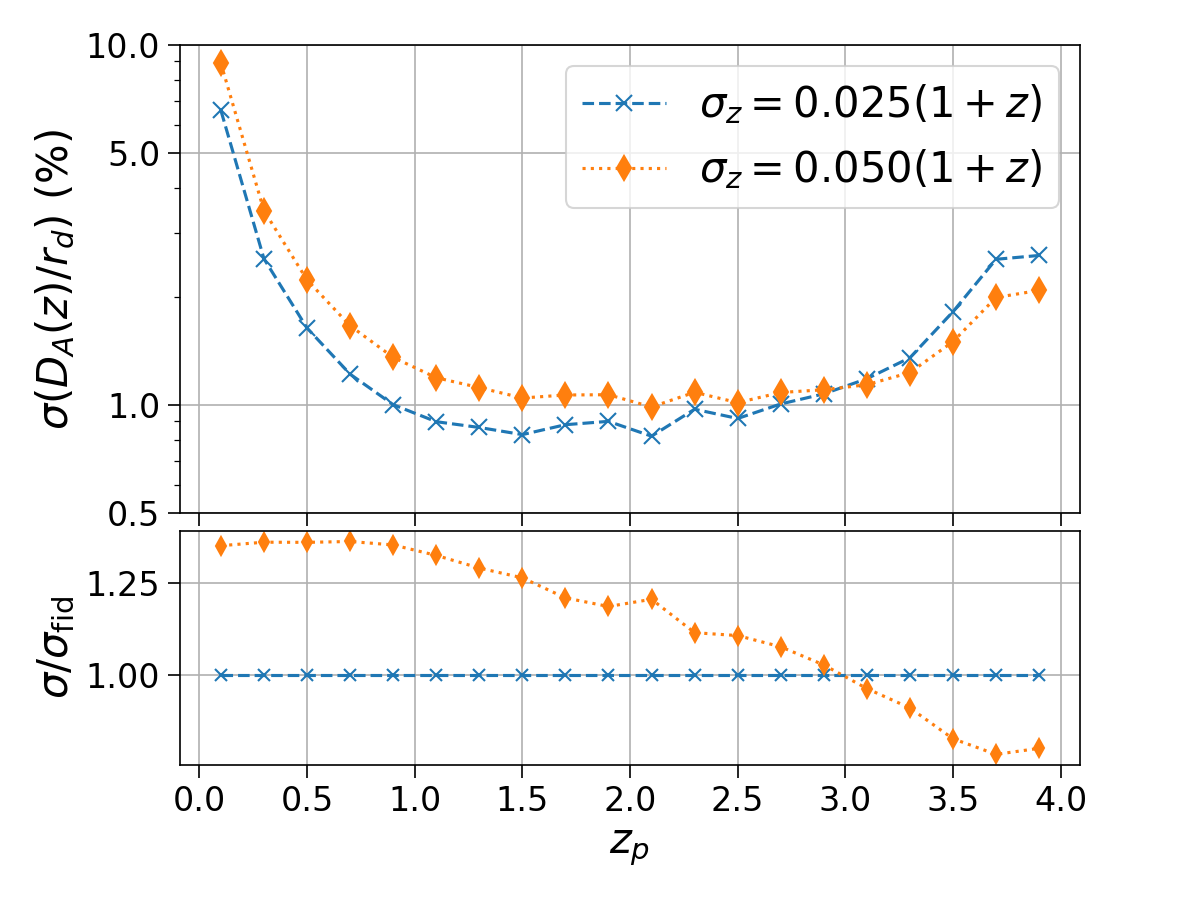}
    \caption{Upper panel: $1\sigma$ constraints on $\DA/\rd$ from the photo-z galaxy power spectrum at the redshift range $0<z<4$ with the bin width $0.2$. We compare the constraints from the two samples with different photo-z errors. Lower panel: The ratio with respect to the fiducial value.}
    \label{fig:photoz_sigma_DA}
\end{figure}

In Fig. \ref{fig:photoz_sigma_DA}, we show the constraints on $\DA/\rd$ from the two photo-z samples as shown in Fig. \ref{fig:nz_phot}. The two photo-z samples have different redshift errors and number densities. For the optimal case without any systematic noise, the constraints from the fiducial sample with the photo-z error $0.025(1+z)$ are $> 20$ per cent tighter than that with the larger photo-z error $0.05(1+z)$ at $z<2$, even though the fiducial number density is $40$ per cent to $60$ per cent lower. At $z>3$, the performance from the other sample surpasses the fiducial one.

%%%%%%%%%%%%%%%%%%
\section{Dependence on the spec-z systematic noise and redshift success rate}
\begin{figure*}
    \centering
    \includegraphics[width=0.9\linewidth]{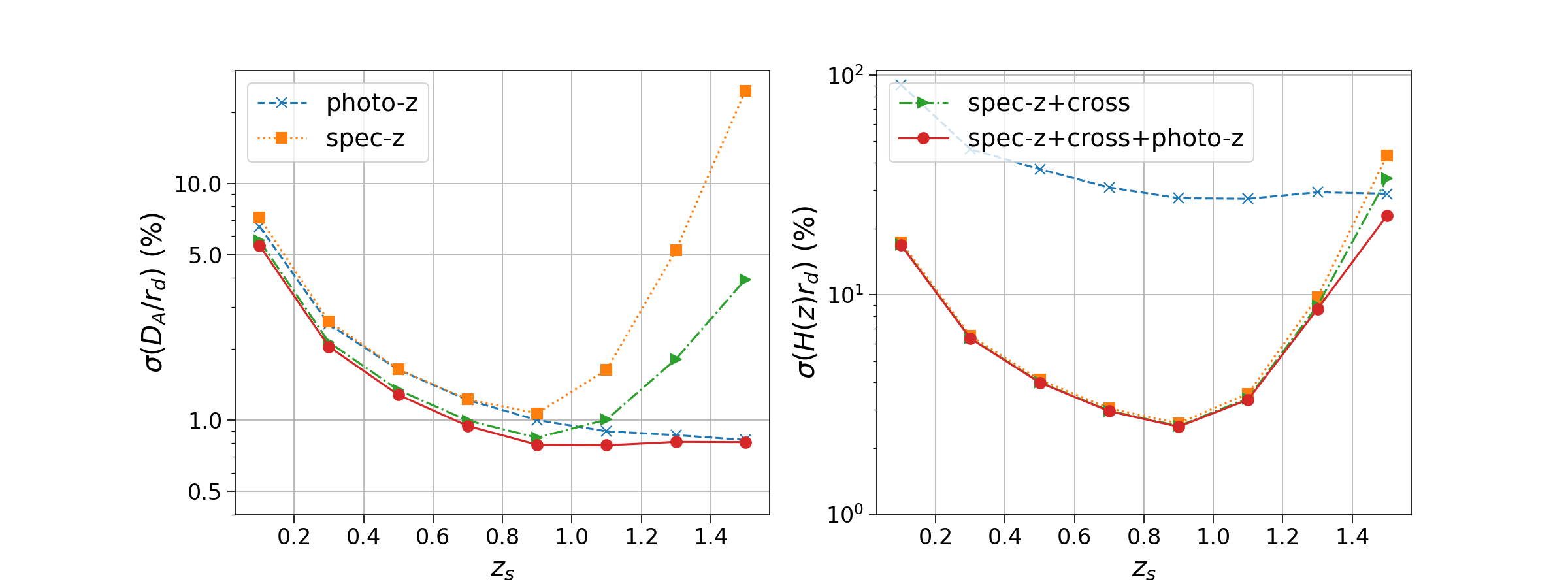}
    \caption{Same as Fig. \ref{fig:sigma_DA_Hz_default} but for the case with the systematic noise $\Psys=2\times 10^3\Mpchcube$ considered in the spec-z power spectrum.}
    \label{fig:sigalperp_specz_cross_Psys2.e3}
\end{figure*}

\begin{figure}
    \centering
\includegraphics[width=0.9\linewidth]{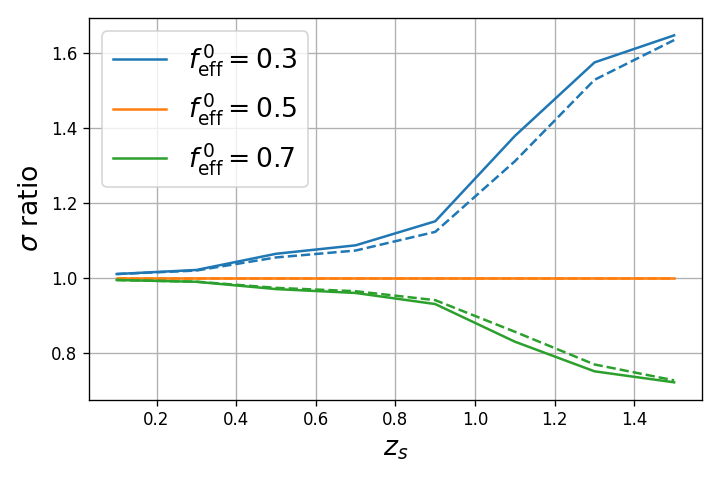}
    \caption{Ratio of the $1\sigma$ constraints on $\DA/\rd$ and $H(z)\rd$ from the spec-z samples with different redshift success rates compared to the fiducial one with $f_{\text{eff}}^0=0.5$. The solid and dashed lines denote for $\DA/\rd$ and $H(z)\rd$, respectively.}
    \label{fig:sigma_diff_f0eff}
\end{figure}

As a case study, Fig. \ref{fig:sigalperp_specz_cross_Psys2.e3} shows the constraints on $\DA/\rd$ and $H(z)\rd$ from the individual tracers and joint analyses. The spec-z power spectrum contains a constant systematic noise $\Psys=2\times 10^3 \Mpchcube$. For $\DA/\rd$, $\Psys$ can largely damp the constraint from the spec-z tracer. The spec-z+cross result is slightly larger than that of Fig. \ref{fig:sigma_DA_Hz_default}, indicating the robustness against the influence of $\Psys$. For $H(z)\rd$, the systematic noise reduces the constraint from the spec-z clustering as well, but less significant than that of $\DA/\rd$. The spec-z constraint on $H(z)\rd$ is still a few times better than that from the photo-z one at $z<1.0$. Therefore, the joint analyses of the spec-z and photo-z clustering do not help the constraint on $H(z)\rd$ even with some amount of the spec-z systematic noise considered.  

Fig. \ref{fig:sigma_diff_f0eff} shows the change of $1\sigma$ error on $\DA/\rd$ and $H(z)\rd$ constrained from the spec-z tracer with different redshift success rates. The relative change on the sigma error is below $10$ per cent at $z<0.8$ even if the galaxy number density is $40$ per cent lower or higher than the default one. Because the default number density is high enough, which is larger than $10^{-3}\hMpccube$ at $z<0.8$, the cosmic variance still dominates even if the number density decreases by such amount. As the number density goes significantly lower at $z>1.0$, the relative change on the number density becomes vital, especially for the case with lower redshift success rate.

% Don't change these lines
\bsp	           % typesetting comment

\label{lastpage}

\end{document}